\documentclass{article}

\usepackage{arxiv}

\usepackage[utf8]{inputenc} 
\usepackage[T1]{fontenc}    
\usepackage{hyperref}       
\usepackage{url}            
\usepackage{booktabs}       
\usepackage{amsfonts}       
\usepackage{nicefrac}       
\usepackage{microtype}      
\usepackage{lipsum}		
\usepackage{graphicx}
\usepackage[round]{natbib}
\usepackage{doi}

\usepackage[thinlines]{easymat} 
\usepackage{amsmath,amsfonts,amssymb} 
\usepackage{indentfirst}
\usepackage{graphicx}
\usepackage{subeqnarray}
\usepackage{mdwlist}
\usepackage[percent]{overpic}
\usepackage{epstopdf}
\usepackage{rotating}
\usepackage{xcolor}
\usepackage{units}

\newcommand\undermat[2]{%
  \makebox[0pt][l]{$\smash{\underbrace{\phantom{%
    \begin{matrix}#2\end{matrix}}}_{\text{$#1$}}}$}#2}

\definecolor{green2}{RGB}{0,128,0}

\makecompactlist{enumerateb}{enumerate}
\makecompactlist{itemizeb}{itemize}
\makecompactlist{descriptionb}{description}

\allowdisplaybreaks 

\def\aujour{\number\day \space \ifcase\month\or
janvier\or fŽvrier\or mars\or avril\or mai\or
juin\or juillet\or aožt\or septembre\or octobre\or
novembre\or dŽcembre\fi \space \number\year}

\def\cH{{\cal H}}

\def\cL{{\cal L}}

\newtheorem{remark}{Remark}

\def\C{{\setbox0=\hbox{$\displaystyle{\rm C}$}
        \hbox{\hbox to0pt{\kern 0.4\wd0\vrule height 0.95\ht0\hss}\box0}}}
\def\Q{{\setbox0=\hbox{$\displaystyle{\rm Q}$}%
    \hbox{\raise 0.2\ht0\hbox to0pt{\kern 0.4\wd0\vrule height
    0.85\ht0\hss}\box0}}} 

\def\cH2{{\cal H}_2} 

\def\cL2{\mathop{\mathcal L}_{2}} 

\def\cRH2{\mathop{\cal R \cal H}_2} 
\def\cRL2{\mathop{\cal R \cal L}_{2}} 

\DeclareMathOperator*{\diag}{diag}

\DeclareMathOperator*{\der}{d}

\makeatletter
\DeclareRobustCommand\sfrac[1]{\@ifnextchar/{\@sfrac{#1}}
                                            {\@sfrac{#1}/}}
\def\@sfrac#1/#2{\leavevmode\kern.1em\raise.5ex
         \hbox{$\m@th\fontsize\sf@size\z@
                           \selectfont#1$}\kern-.1em
         /\kern-.15em\lower.25ex
          \hbox{$\m@th\fontsize\sf@size\z@
                            \selectfont#2$}}
\makeatother
\title{Model-free LQR based PID controller for trajectory tracking of 2-DoF helicopter: comparison and experimental results}

\author{ \href{https://orcid.org/}{\includegraphics[scale=0.06]{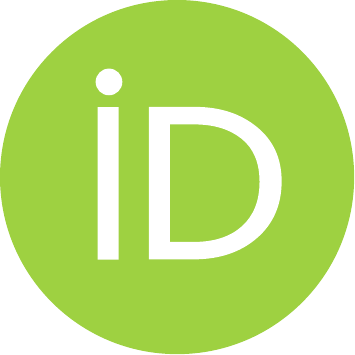}\hspace{1mm}Nouha Rouis$^1$$^2$}, 
\href{https://orcid.org/0000-0002-2576-5515}{\includegraphics[scale=0.06]{orcid.pdf}\hspace{1mm}Ibrahima~N'Doye$^1$}, 
\href{https://orcid.org/0000-0001-5944-0121}{\includegraphics[scale=0.06]{orcid.pdf}\hspace{1mm}Taous-Meriem~Laleg-Kirati$^1$}
\thanks{This work has been supported by the King Abdullah University of Science and Technology (KAUST), Base Research Fund (BAS/1/1627-01-01) to Taous Meriem Laleg.} \\
$^1$Computer, Electrical and Mathematical Sciences and Engineering Division (CEMSE)\\
King Abdullah University of Science and Technology (KAUST)\\
Thuwal 23955-6900, Saudi Arabia \\
	\texttt{ibrahima.ndoye@kaust.edu.sa; taousmeriem.laleg@kaust.edu.sa} \\
$^2$RISC Laboratory, National School of Engineers of Tunis\\ 
University of Tunis El Manar\\
\texttt{nouha.rouis@kaust.edu.sa}\\
}

\renewcommand{\shorttitle}{\textit{arXiv} Template}

\hypersetup{
pdftitle={A template for the arxiv style},
pdfsubject={q-bio.NC, q-bio.QM},
pdfauthor={David S.~Hippocampus, Elias D.~Striatum},
pdfkeywords={First keyword, Second keyword, More},
}

\begin{document}
\maketitle

\begin{abstract}
This paper studies the performance of a model-free LQR based PID (i-LQR-PID) controller designed for tracking control problem of a 2-DoF laboratory helicopter. The control problem addressed in 2-DoF helicopter system aims to track the desired pitch and yaw axes trajectories despite disturbed operating conditions. In addition to the unpredictable variations, the 2-DoF helicopter dynamic is highly nonlinear with having strong cross-couplings in their models as well as being open loop unstable system. Thus, we propose a model-free LQR based PID control strategy in order to achieve better trajectory tracking control objectives. Robustness tests are performed experimentally to show the effectiveness of the model-free control.  
\end{abstract}


\section{Introduction}
Design of  control strategies for helicopters  has attracted the research community due to its wide range of  civil and military applications \cite{ZhZ:11} and it is still an active research topic with several challenges including the presence of  high nonlinearities and model uncertainties \cite{HAH:12}. Both stabilization and trajectory tracking problems for helicopters systems  have been studied in the literature.  For instance, a combined feed forward action and saturation feedback was proposed in \cite{MaN:07}. A robust Linear Quadratic Regulator (LQR) was introduced for attitude control of 3-DoF helicopter in \cite{LLZ:13}. Moreover, a backstepping based approach \cite{RVM:11} and an adaptive LQR using Model Reference Adaptive Control (MRAC) scheme \cite{SuE:16} have been proposed to solve tracking problems in unmanned helicopters.   However, there is still a need for robustness enhancement especially under aggressive turbulence effects.

In this study, a model-free LQR-proportional-integral-derivative controller called intelligent LQR-proportional-integral-derivative (i-LQR-PID) controller is introduced as an alternative robust control strategy to the LQR-PID control. The conventional PID controller is one of the most used controller in industry for closed control-loops thanks to its simplicity in real time implementation. The design and tuning of such control algorithm has been widely covered and is still  an active field of research especially for industrial plants subject to external disturbances \cite{AsH:01,HLHN:99,Ast1993}. Indeed, due to the significant variations in the amplitude vibration affected by the external disturbances, satisfactory performance covering the total range of disturbances is difficult to reach with a conventional PID without an external compensation. Thereafter, it is desirable to design robust control strategies without additional computational effort. 

To consider modeling errors, system uncertainties, disturbances and actuator faults when designing a controller, a model-free control (MFC) algorithm has been proposed in \cite{FlJ:09}, \cite{FlJ:13}. The main feature of this approach consists in updating continuously the input-output behavior using an ultra-local-model.  To improve the performance and robustness of conventional controllers  with less time and effort expenditure,  MFC has been successfully combined to some controllers  such as PID controller and more recently the  LQR  providing the so-called intelligent PID controller (i-PID) and intelligent LQR (i-LQR) \cite{AsH:06}. 


MFC in general, i-PID and i-LQR controllers in particular have been considered in several applications  and their performance have been studied in both simulation and experiments. Examples of such applications include shape memory alloys \cite{GJDBCC:09}, DC/DC converters \cite{MJFSC:10}, active magnetic bearing \cite{MJFRB:09}, two-dimensional planar manipulator \cite{MaH:13}, agricultural greenhouse \cite{LBPF:15}, quadrotor vehicle and aerospace \cite{MAFM:13,YDNRH:16,AFJ:17,MAFGM:17}, automotive engine \cite{CAFMV:09}, mechanical system \cite{ViH:12} and Qball-X4 quadrotor vehicle \cite{YDNRH:14}, \cite{YDNRH:16}. 
  
In this paper,  an intelligent LQR based PID (i-LQR-PID) controller is designed  for reference trajectory tracking of  the pitch and yaw angles in a helicopter system.  The performance of the i-LQR-PID is evaluated by comparison to LQR-PID controller through experiments. Moreover, robustness analysis is performed and validated experimentally with respect to nominal tracking, exogenous disturbance, parameter uncertainty and wind disturbances through experiments. 


\section{System description}\label{sec-description}
\begin{figure}[ht]
\centering
\begin{overpic}[scale=0.98]{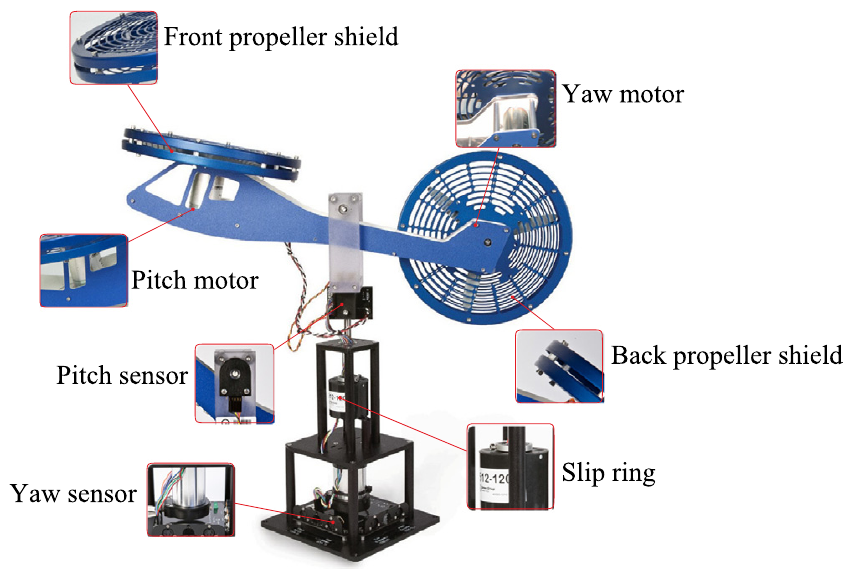}
\end{overpic} 
\caption{2-DoF Helicopter Workstation \cite{Qua:10}}
\label{fig_workstation}
\end{figure}

The Quanser 2-DoF laboratory helicopter has been studied in this paper. As illustrated in Fig. \ref{fig_workstation}, the system consists of  a helicopter body on a fixed base with two propellers.  DC motors drive these propellers which control both the pitch and yaw angles of the helicopter.

\vspace{0.25cm}

\begin{figure}[ht]
\centering
\begin{overpic}[scale=0.98]{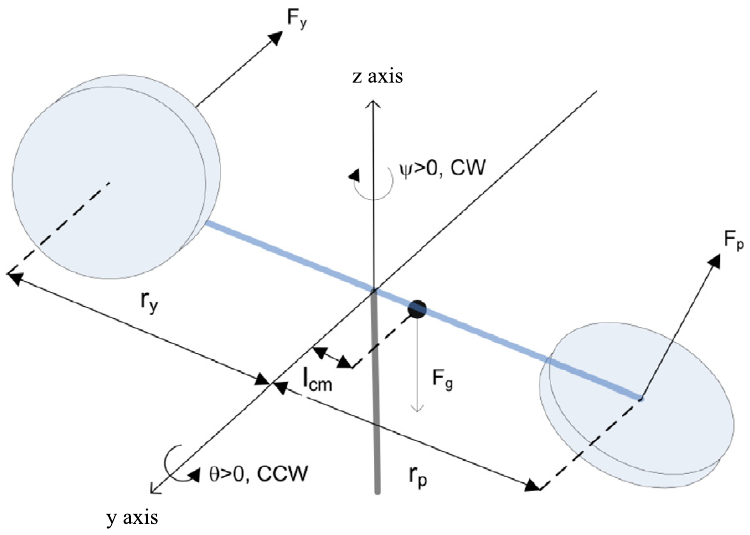}
\end{overpic} 
\caption{Free body diagram of 2-DoF laboratory}
\label{fig_diagram}
\end{figure}

\vspace{0.25cm}
Fig. \ref{fig_diagram} shows the free body diagram of 2-DoF laboratory helicopter. There are two degrees of freedom that are given by the  motion around the $z$ axis (yaw), represented by angle $\Psi$, and the rotation around the $y$ axis (pitch), represented by the angle $\theta$. The control inputs for the system are given by the  voltages to the DC motors.
\vspace{0.25cm}
\vspace{0.25cm}

Using the kinematic diagram of the 2-DoF helicopter testbed and Euler Lagrangian energy based approach, the state space representation model governing the dynamics of the 2-DoF helicopter system is given in \eqref{eq-1} (see \cite{Qua:10}).
\begin{table*}[ht]
\begin{center}
\line(1,0){480}
\end{center}
\begin{align}\label{eq-1} 
&\!\!\begin{bmatrix}\dot{\theta}\\ \dot{\Psi} \\ \ddot{\theta} \\ \ddot{\Psi} \\ \dot{I}_{\theta}\\ \dot{I}_{\Psi} \end{bmatrix}\!\!=
\!\!\begin{bmatrix} 0 &0 & 1 &0 &0 &0 \\ 
0 & 0 & 0 &1 &0 &0 \\
0 & 0 & \frac{-B_p}{J_{eq,p}+m_h\ell^2} &0  &0 &0 \\
0 & 0 & 0 &\frac{-B_y}{J_{eq,y}+m_h\ell^2} & 0&0 \\
1 & 0 & 0 &0 &0 &0 \\
0 & 1 & 0 &0 &0 &0 \end{bmatrix}\begin{bmatrix}\theta\\ \Psi \\ \dot{\theta} \\ \dot{\Psi} \\ I_{\theta}\\ I_{\Psi}\end{bmatrix} 
\!+\!\begin{bmatrix} 0 & 0 \\ 0 & 0 \\ \frac{K_{pp}}{J_{eq,p}+m_h\ell^2}  & \frac{K_{py}}{J_{eq,p}+m_h\ell^2}  \\\frac{K_{yp}}{J_{eq,y}+m_h\ell^2}  & \frac{K_{yy}}{J_{eq,y}+m_h\ell^2}   \\0 & 0 \\0 & 0 \end{bmatrix}\!\begin{bmatrix}u_p\\ u_y\end{bmatrix}, \notag \\&
Y=\begin{bmatrix} 1 & 0 & 0 & 0 & 0 & 0\\ 0 & 1 & 0 & 0 & 0 & 0 \end{bmatrix}\begin{bmatrix}\theta\\ \Psi \\ \dot{\theta} \\ \dot{\Psi} \\ I_{\theta}\\ I_{\Psi}\end{bmatrix},\qquad  I_{\theta}=\int (\theta-\theta_d)\der t, \qquad I_{\Psi}= \int (\Psi -\Psi _d)\der t.
\end{align}
\begin{center}
\line(1,0){480}
\end{center}
\end{table*}

The nominal plant parameters of the 2-DoF helicopter system are given in Table \ref{tab_param}.
\begin{table}[ht]
\begin{center}
\caption{Nominal parameters of 2-DoF Helicopter model.}\label{tab_param}
\begin{tabular}{p{0.07\textwidth}p{0.07\textwidth}p{0.08\textwidth}p{0.2\textwidth}}
\hline
Symbol & Unit & Value &Description \\
\hline
$B_p$&$N/V$& $0.8$ &Equivalent viscous damping about pitch axis\\
$B_y$&$N/V$& $0.318$ &Equivalent viscous damping about yaw axis.\\
$J_{eq,p}$&$kg.m^2$& $0.0384$ &Total moment of inertia about pitch axis.\\
$J_{eq,y}$&$kg.m^2$& $0.0432$ &Total moment of inertia about yaw axis.\\
$m_h$&$kg$& $1.3872$  &Total moving mass of the helicopter\\
$\ell$ &$m$ & $0.186$   &Length along helicopter body from pitch axis.\\
$K_{pp}$&$N.m/V$& $0.204$ &Thrust force constant of pitch motor/propeller.\\
$K_{py}$&$N.m/V$& $0.0068$  &Thrust torque constant acting on pitch axis from yaw motor/propeller.\\
$K_{yp}$& N.m/V& $0.0219$ &Thrust torque constant acting on yaw axis from pitch motor/propeller.\\
$K_{yy}$&$N.m/V$& $0.072$ &Thrust torque constant acting on yaw axis from yaw motor/propeller.\\
$u_p$ &$V$& $\pm24$ &Pitch motor voltage.\\
$u_y$ &$V$& $\pm15$  &Yaw motor voltage.\\
\hline
\end{tabular}
\end{center}
\end{table}

The control objective of this study is to track a desired reference trajectory for both pitch and yaw angles using LQR based PID and intelligent LQR based PID  controllers.

\section{LQR based PID tracking control}\label{sec-LQR}
Linear-quadratic-regulator (LQR) is an optimal control strategy \cite{Nai:02}, \cite{Cho:14} which has been widely used in various applications. LQR design is based on the selection of a state feedback gain $K$ such that the cost function or performance index $J$ is minimized. This ensures that the gain selection is optimal for the cost function specified.  The inherent robustness, stability and optimality between the control input and speed of response make LQR as the most preferred optimal controller in aerospace applications.

Consider the following state space representation of the augmented 2-DoF helicopter system 
\begin{equation}\label{lin}
\left \{
\begin{array}{l}
\dot{x}(t)=Ax(t)+Bu(t),\\
y(t)=Cx(t)
\end{array}
\right.
\end{equation}
The PID control gains are computed using the LQR scheme where $x^T=\begin{bmatrix}\theta & \Psi & \dot{\theta} & \dot{\Psi} & I_{\theta}& I_{\Psi}\end{bmatrix}$.

Using the state feedback control law,
\begin{equation}\label{eq-lin1}
u(t)=-Kx(t),
\end{equation}
The LQR controller proposes an optimal control strategy  that minimizes the following cost function.
\begin{equation}\label{eq-lin2}
J(u^*)=\int_0^\infty[x^T(t)Qx(t)+u^T(t)Ru(t)]\der t, 
\end{equation}
where  $Q$ and $R$ are some weighting matrices  that penalize the state variables  and  the inputs. 
The control input using the LQR approach is given by
\begin{equation}\label{eq-lin3}
K=R^{-1}B^TP.
\end{equation}
The transformation matrix $P$ is the solution of the following Algebraic Riccati Equation
\begin{equation}\label{eq-lin4}
A^TP+PA+Q-PBR^{-1}B^TP=0.
\end{equation}
The LQR usually offers good robustness properties  but its performance deteriorates under severe uncertainties which are for instance due to actuator failure or structural damage \cite{OLAQ:09}. To enhance the trajectory tracking performance of the LQR, an intelligent LQR based PID controller framework is proposed in the next section. 

\section{Intelligent LQR based PID tracking control}\label{sec-iPID}
The main idea behind a model-free technique is to update continuously the behavior input/output using an ultra-local model. Therefore, the resulting control strategy is robust with respect to un-modeled dynamics and uncertainties of the system. Model-free control has been successfully applied in several applications (\cite{FlJ:13,FlJ:14,AFJ:17,CAFMV:09,MAFM:13,VACFM:09,ViH:12,YDNRH:16,MAFGM:17}). 

The following  ultra-local model is usually used in MFC
\begin{equation}\label{Model-free1}
y^{(\nu)}(t) = F(t) +\alpha u(t)
\end{equation}
where $u$ is the control input, $y$ is the  output, $F$  refers to all unknown model of the system, disturbances  and uncertainties which is estimated from available input/output measurements, $\nu$ is the differentiation order, $\alpha$ is a constant which is chosen such that $\alpha u$ and $y^{(\nu)}$ are of the same order of magnitude.
\begin{remark} 
The proper value of $\nu$ has to be chosen. This value usually depends on prior knowledge on the system and its relative degree. It may also depend on the nature of the control input. Indeed, studies have shown that with an intelligent Proportional-Integral-Derivative (iPID) and iPD controllers the value $\nu\!=\!2$ is suitable while for iPI and iP a value $\nu\!=\!1$ is more suitable \cite{YDNRH:16}.
\end{remark}
In this paper, the proper value of $\nu$ is setting to $2$, due to the fact that the relative degree of system \eqref{eq-1} is $2$ and i-PID controller is also used in the 2-DoF helicopter system~\eqref{eq-1}. 

Setting $\nu=2$ in equation \eqref{Model-free1}, we have
\begin{equation}\label{Model-free2}
\ddot{y}(t)= F(t) +\alpha u(t)
\end{equation}
Hence, the corresponding i-PID can be written as
{\small
\begin{equation}\label{Model-free3}
u(t)= -\displaystyle\frac{(\widehat{F}(t)-\ddot{y}^d(t)+K_Pe(t)+K_D\dot{e}(t)+K_I  \int_0^t e(\tau) \der \tau)}{\alpha}
\end{equation}}
where 
\begin{itemize}
  \item $y^d$ is the desired reference trajectory,
  \item $\widehat{F}(t)$ is the estimate of $F(t)$ which is described as follows,
  \begin{equation}\label{Model-free2bis}
\widehat{F}(t)=\widehat{\ddot{y}}(t)-\alpha u(t),
\end{equation}
 \item $\widehat{\ddot{y}}$ is the estimate of  $\ddot{y}$,
  \item $e(t)=y(t)-y^d(t)$ is the trajectory tracking error,
  \item $K_P$, $K_I$ and $K_D$ are the PID gains.
\end{itemize}
Once $\widehat{\ddot{y}}$ is obtained, the loop is closed by \eqref{Model-free3} and the modified expression of \eqref{Model-free2} is given as \eqref{Model-free2bis}.

Substituting  equation \eqref{Model-free2bis} in \eqref{Model-free3}, adding and subtracting the derivate of the controlled output $\ddot{y}$ yields
\begin{equation}\label{Model-free4}
 \ddot{e}(t)+K_Pe(t)+K_D\dot{e}(t)+K_I  \int_0^t e(\tau) \der \tau=e_F(t),
\end{equation}
where $e_F(t)\!=\!\ddot{y}(t)-\widehat{\ddot{y}}(t)\!=\!F(t)-\widehat{F}(t)$. Subsequently, with a good estimate $\widehat{F}(t)$ of $F(t)$ {\it i.e} $e_F(t)\!=\!F(t)-\widehat{F}(t)\simeq0$, equation \eqref{Model-free4} yields
\begin{equation}\label{Model-free4bis}
\ddot{e}(t)+K_Pe(t)+K_D\dot{e}(t)+K_I  \int_0^t e(\tau) \der \tau\simeq0,
\end{equation} 
which ensures an excellent tracking of the reference trajectory. This tracking is moreover quite robust with respect to uncertainties and disturbances which can be important in the 2-DoF helicopter stabilization setting such as considered here. This robustness feature is explained by the fact that $F$ includes all the effects of unmodeled dynamics and disturbances, without trying to distinguish between its different components. Furthermore, the approximation of PID design parameters becomes therefore quite straightforward. This is a major benefit when compared to ``classic" PIDs.
\begin{figure}[ht]
\vspace{0.25cm}
        \centering
      \begin{overpic}[scale=0.26]{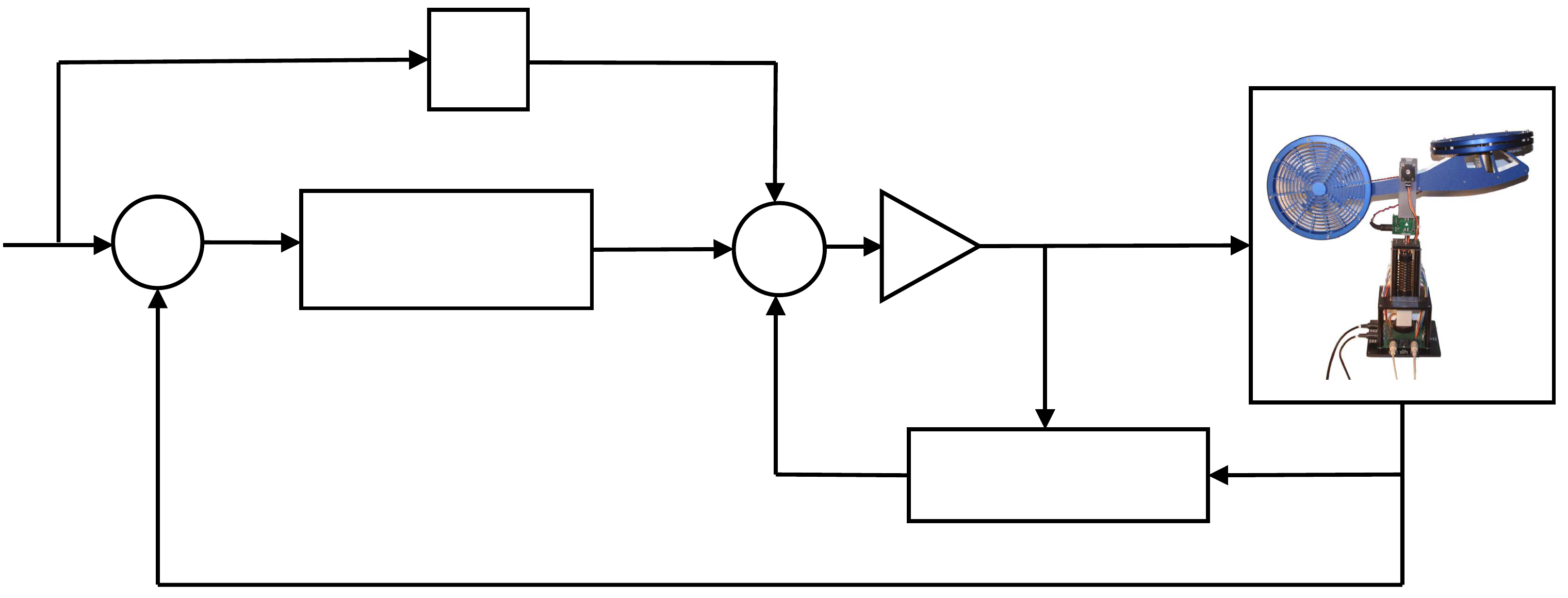}
   \put(28,33.5){\scriptsize $\frac{d^2}{\der\!t^2 }$}
 \put(-5,28){\scriptsize $\begin{bmatrix}\theta_d \\ \Psi_d\end{bmatrix}$}
 \put(50,32){\scriptsize $\begin{bmatrix}\ddot{\theta}_d \\ \ddot{\Psi}_d\end{bmatrix}$}
 \put(71,27){\scriptsize $\begin{bmatrix}u_p \\ u_y\end{bmatrix}$}
  \put(23,22){\scriptsize LQR$_{\mbox{PID}}$}
    \put(56.5,21.65){\scriptsize $\frac{1}{\alpha}$}
     \put(90,6){\scriptsize $\begin{bmatrix}\theta \\ \Psi\end{bmatrix}$}
     \put(50,14){\scriptsize $F$}
      \put(61,8){\scriptsize Ultra-local}
      \put(64,5.5){\scriptsize Model}
      \put(49,23){\scriptsize +}
      \put(48,21.5){\scriptsize -}
      \put(49,20){\scriptsize -}
      \put(8,21.5){\scriptsize +}
      \put(10,20.5){\scriptsize -}
            \end{overpic}  \vspace{-6pt}
              \caption{Schematic diagram of LQR-PID model-free control of 2-DoF helicopter.}\label{diag-heli}           
\end{figure}
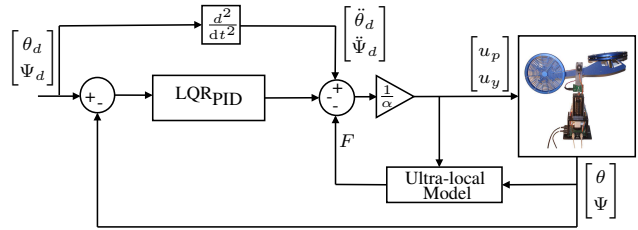

\section{Simulation and experimental results: Comparison and discussions}\label{sec-Exp-results}
\begin{figure}[ht]
\vspace{0.5cm}
\centering
\begin{overpic}[scale=0.34]{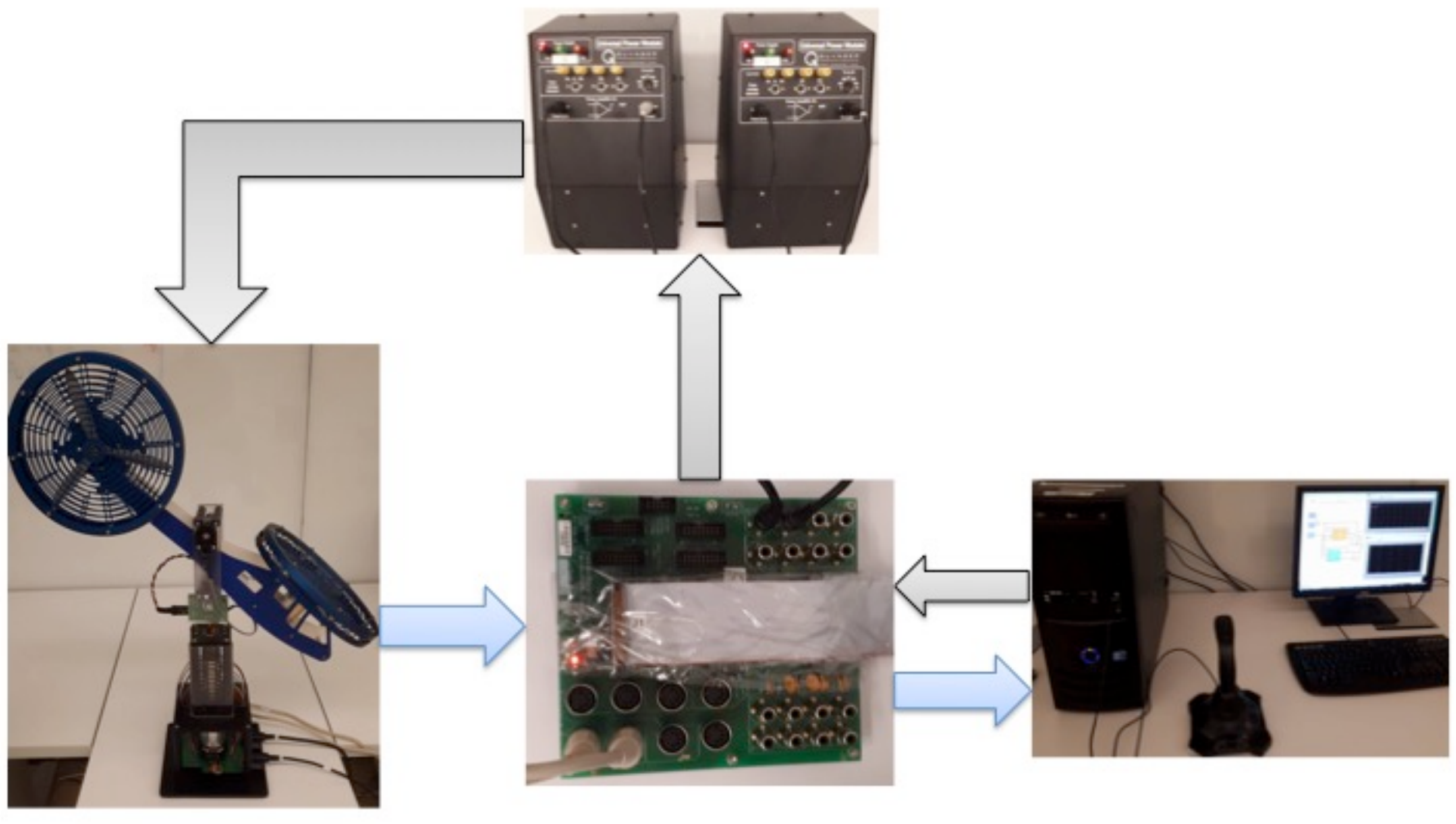}
\put(69,1){\scriptsize \textcolor{black}{Real-time control software}}
\put(39,-1){\scriptsize \textcolor{black}{Data Acquisition}}
\put(4,-2){\scriptsize \textcolor{black}{Helicopter plant}}
\put(39,56){\scriptsize \textcolor{black}{Power amplifier}}
\put(6,40){\scriptsize \textcolor{black}{Control signal}}
\put(31,5){\scriptsize \begin{rotate}{90}  \textcolor{black}{Encoder data}\end{rotate}}
\put(61.5,18){\scriptsize \textcolor{black}{Control}}
\put(62,12){\scriptsize \textcolor{black}{signal}}
\end{overpic} \vspace{-2pt}
\caption{2-DoF helicopter experimental system}
\label{fig_Exp}
\end{figure}
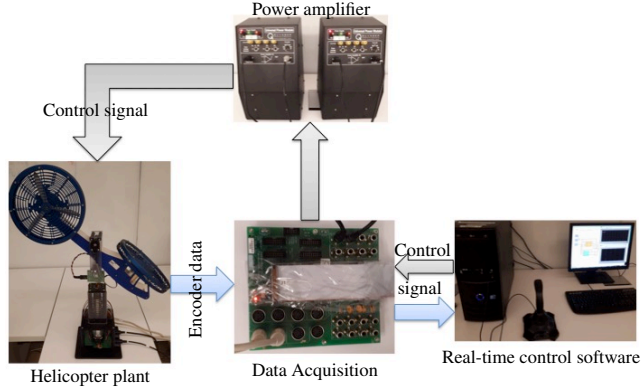

Fig. \ref{fig_Exp} illustrates the hardware in the loop experimental system of a 2-DoF helicopter. Its consists of helicopter plant, data acquisition board, computer and power amplifier. This system is a real-time control experimental system with the support of real-time software QUARC. The blue arrows represent the acquisition of current states, while the gray arrows are actions of control solution. Integrated encoders measure the pitch and yaw angles. The controller is supervised by commands from a Matlab/Simulink based human machine interface.\\

\vspace{0.25cm}

To evaluate the effectiveness of the intelligent LQR based PID (i-LQR-PID) over the LQR based PID (LQR-PID) controller in tracking the desired trajectories despite disturbed operating conditions, the pitch and yaw motors are driven simultaneously by model-free control laws. LQR-PID controller is mainly suitable for linear systems or linearized systems around a specific operating point. However, the tuning of the LQR-PID control law proves to be challenging and highly dependent on the model of the 2-DoF helicopter system. In this work, the 2-DoF helicopter system is decomposed into two-SISO subsystems that are linked to each other and two model-free controls are designed to control simultaneously the pitch and yaw angles. The gain $\alpha$ for each controller is determined empirically. Based on the state representation given in \eqref{eq-1}, $Q$ and $R$ matrices that achieve an optimal control are given as follows:
\begin{equation}\label{equationLQR1}
Q=\diag\begin{bmatrix} 200 & 150 & 100 & 200 & 50 & 50 \end{bmatrix},
\end{equation}
\begin{equation}\label{equationLQR2}
R=\diag\begin{bmatrix} 1 & 1 \end{bmatrix} \, \mbox{with}\ \ \alpha_{\mbox{\scriptsize pitch}}\!=\!1.3 \ \  \mbox{and}\ \ \alpha_{\mbox{\scriptsize yaw}}\!=\!0.43.
\end{equation}
Using the above weighting matrices, the LQR state feedback controller gain $K$ for the system is given as
\begin{equation*}
K\!=\!\left[\!\!\begin{array}{rr|rr|rr} 18.9 & 1.98 & 7.48 & 1.53 & 7.03 & 0.77\\ 
\undermat{K_P}{-2.22 & 19.4} &\undermat{K_D}{-0.45 & 11.9} & \undermat{K_I}{-0.77 &7.03}\!\!\end{array}\right]\!\!.
\end{equation*}

\vspace{0.5cm}

\subsection{Trajectory tracking}\label{sec-TT}
\begin{figure*}[ht]
\vspace{0.25cm}
   \begin{minipage}[c]{0.475\linewidth}  
        \centering
      \begin{overpic}[scale=0.32]{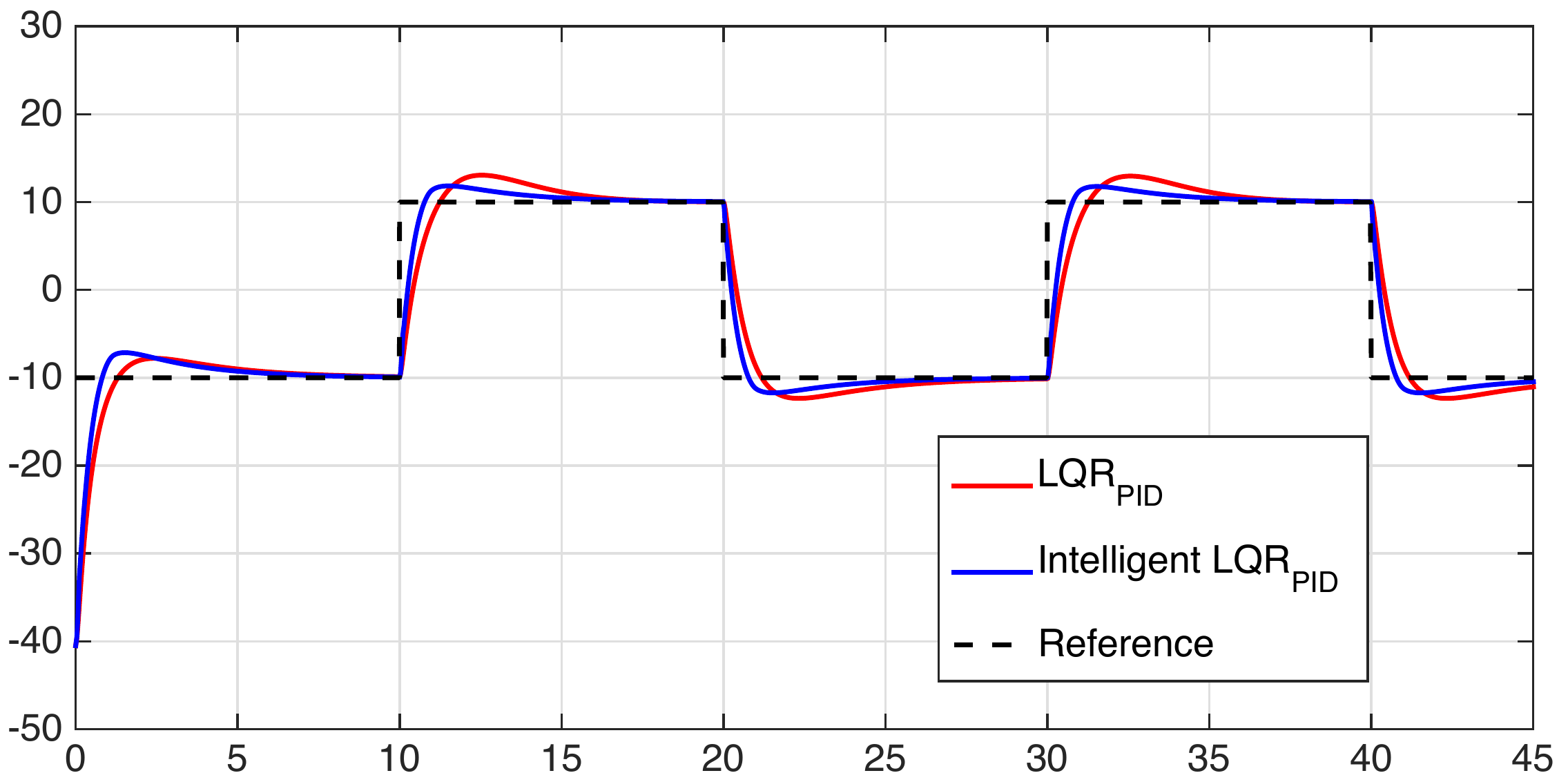}
      \put(-1.5,16){\scriptsize \begin{rotate}{90} {Pitch angle $(\deg)$}\end{rotate}}
 \put(42,-4){Time [sec]}
            \end{overpic}  \vspace{2pt}
              \caption{Simulated pitch tracking responses under pitch reference step.}\label{fig-simP}        
       \end{minipage}\hfill 
         \begin{minipage}[c]{0.475\linewidth}
      \begin{overpic}[scale=0.32]{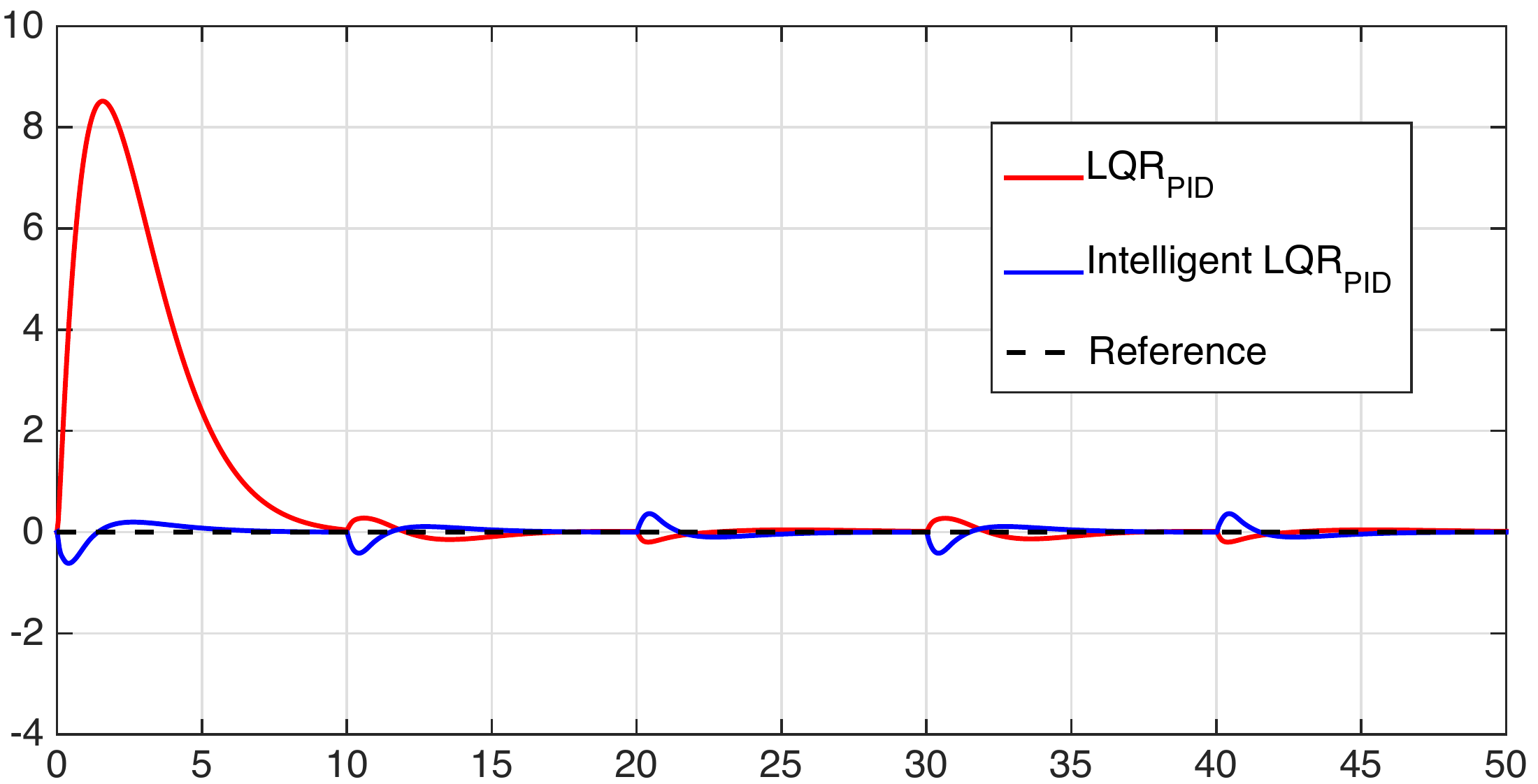}
      \put(-1.5,16){\scriptsize \begin{rotate}{90} {Yaw angle $(\deg)$}\end{rotate}}
 \put(42,-4){ Time [sec]}
            \end{overpic}  \vspace{2pt}
              \caption{Simulated yaw tracking responses under pitch reference step.}\label{fig-simY}
   \end{minipage} 
   \vspace{0.25cm}     
\end{figure*}

The performance analysis of 2-DoF helicopter system has been carried out for pitch and yaw angles through simulations and experiments. The initial position of the helicopter is set to $\theta\!=\!-40.5^{\circ}$ and $\Psi\!=\!0^{\circ}$. To assess the control tracking performance of the intelligent LQR based PID (i-LQR-PID) controller under nominal conditions, a square trajectory with an amplitude of $\pm10^{\circ}$ is given as a reference signal. Figs. \ref{fig-simP} and \ref{fig-simY}, which show the simulated pitch and yaw tracking responses of the pitch reference step scheme for both i-LQR-PID and LQR-PID provide satisfactory results and highlight that the pitch and yaw angles from the i-LQR-PID settles faster at the reference value with a short convergence time of $4$ seconds.

\begin{figure*}[ht]
\vspace{0.25cm}
   \begin{minipage}[c]{0.475\linewidth}  
        \centering
      \begin{overpic}[scale=0.32]{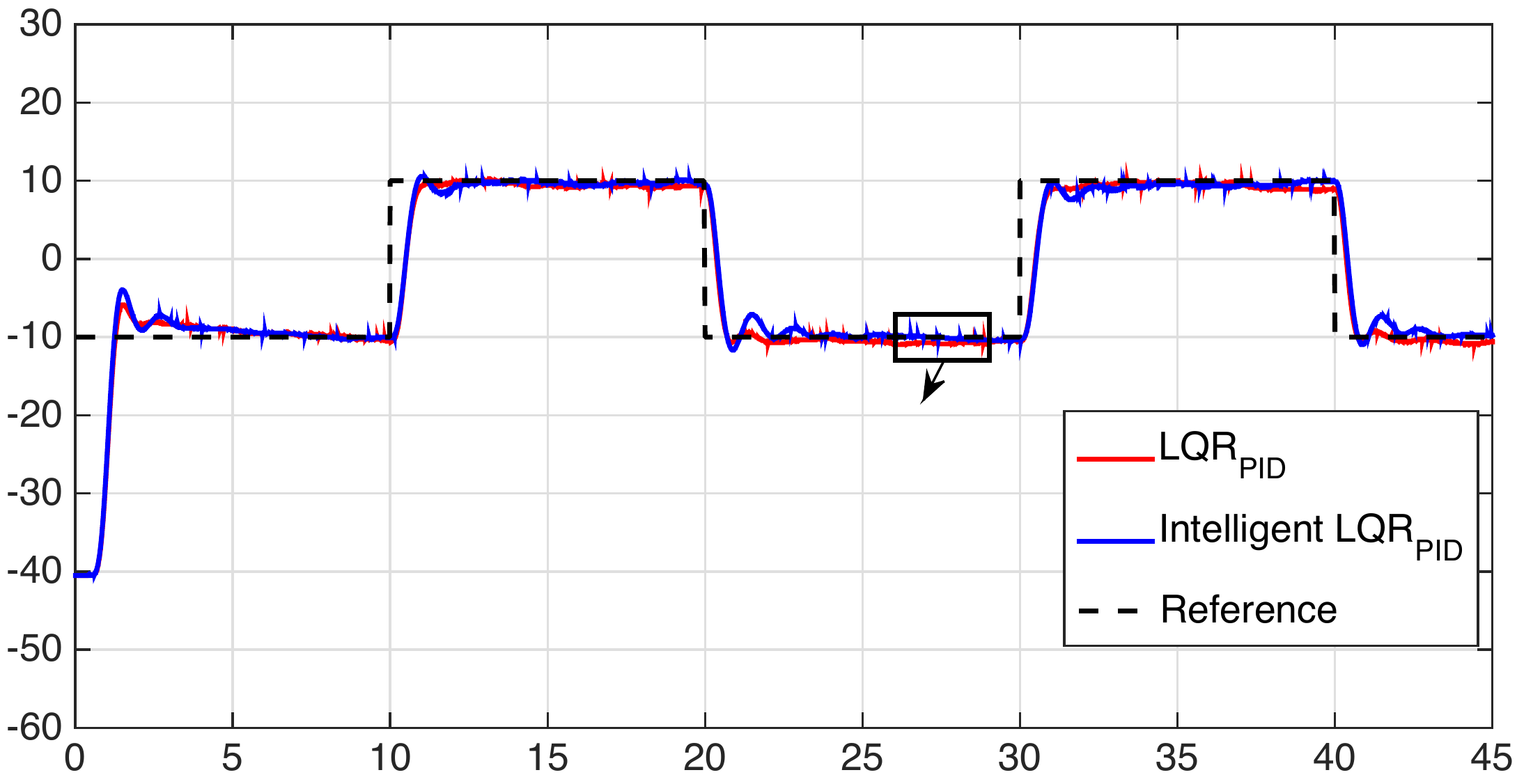}
      \put(-1.5,16){\scriptsize \begin{rotate}{90} {Pitch angle $(\deg)$}\end{rotate}}
     \put(20,5){\begin{overpic}[scale=0.13]{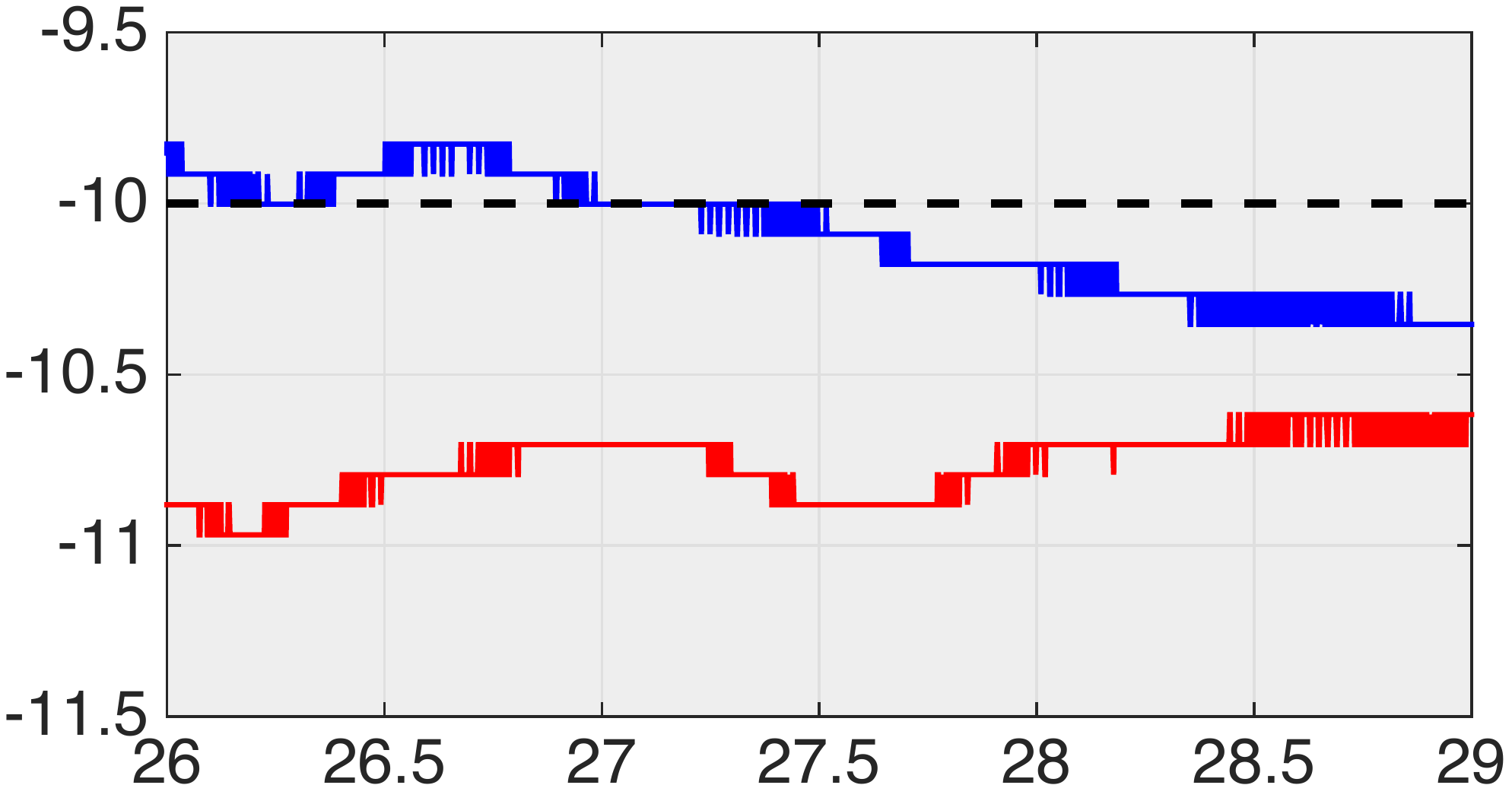}
   \end{overpic}}
 \put(42,-4){Time [sec]}
            \end{overpic}  \vspace{2pt}
     \caption{Experimental pitch tracking responses under pitch reference step.}\label{fig-trackP}     
       \end{minipage}\hfill 
         \begin{minipage}[c]{0.475\linewidth}
      \begin{overpic}[scale=0.32]{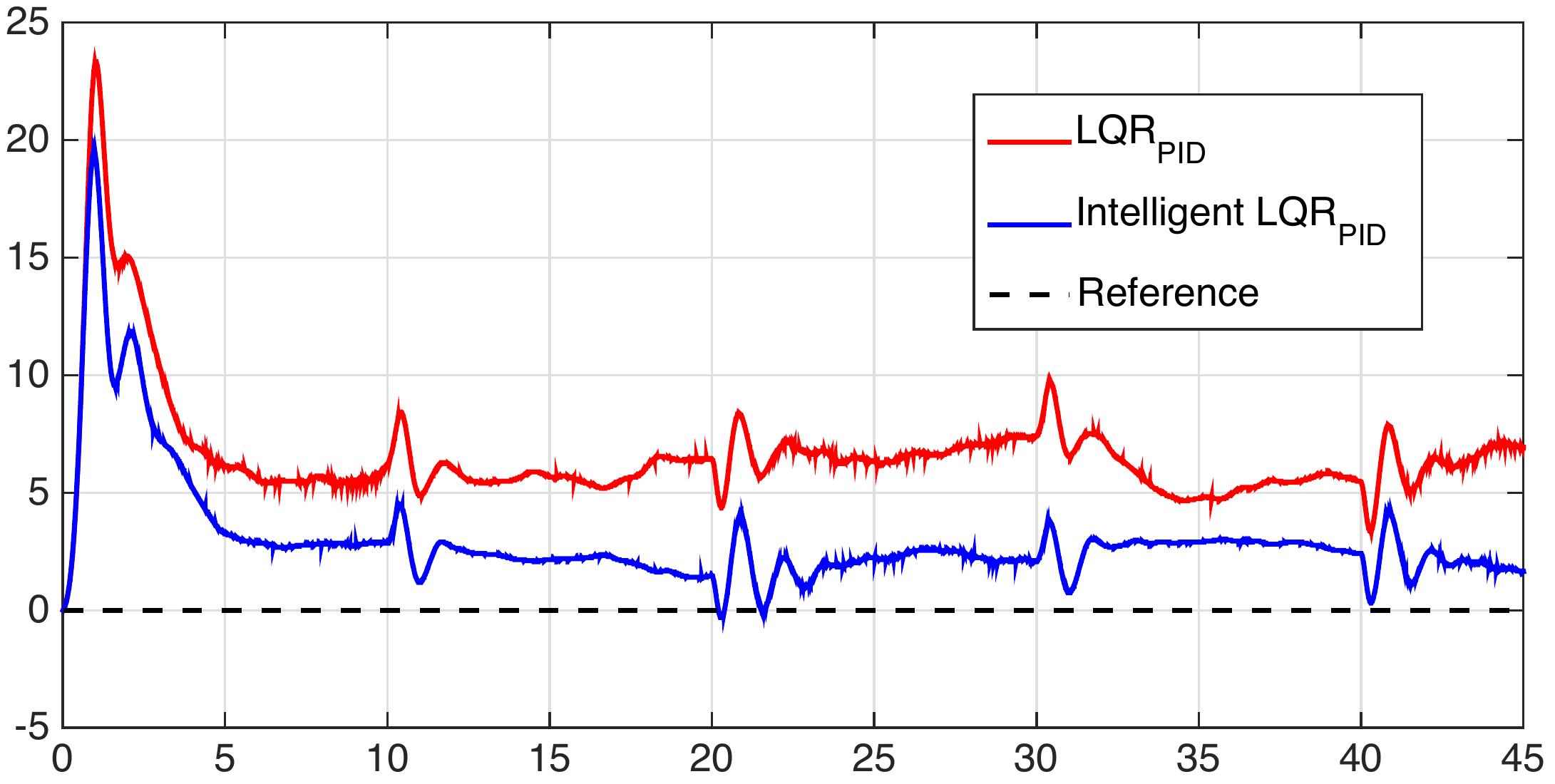}
      \put(-1.5,16){\scriptsize \begin{rotate}{90} {Yaw angle $(\deg)$}\end{rotate}}
 \put(42,-4){Time [sec]}
            \end{overpic}  \vspace{2pt}
              \caption{ Experimental yaw tracking responses under pitch reference step.}\label{fig-trackY}
   \end{minipage} 
   \vspace{0.25cm}     
\end{figure*}

The experimental pitch and yaw tracking responses under pitch reference step are shown in Figs. \ref{fig-trackP} and \ref{fig-trackY} respectively. In Fig. \ref{fig-trackP}, the zoomed view response of i-LQR-PID controller for the pitch angle over the time interval $[26-29]$ seconds gives a better trajectory tracking response. The yaw angle for both controllers shows an offset from the desired trajectory. However, the i-LQR-PID controller reduces the yaw offset angle and presents the smallest tracking error than the LQR-PID controller. The tracking trajectory performance is also evaluated by Root Mean Square (RMS), Standard Deviation (STD) and Mean for both pitch and yaw angles. The performances indices are given in Table \ref{sample-tableP} and Table \ref{sample-tableY} for both pitch and yaw angles respectively. The RMS represents the tracking error between the desired trajectory and the output, STD estimates the central tendency of the distribution of the output and the mean is used to indicate the spread of control results and evaluate the precision of the system. From Tables \ref{sample-tableP} and \ref{sample-tableY}, it can be clearly seen that i-LQR-PID controller achieves better trajectory tracking performances than LQR-PID.

\begin{table}[ht]
\vskip 0.15in
\begin{small}
\caption{\textbf{Pitch} tracking errors}
\begin{center}
\begin{tabular}{| c || c c  c  c |}
  \hline
~Controller~  & ~~ RMS~~  &  STD  & Mean &~~~~Interval~~~~   \\ \hline
~~LQR$_{\mbox{\scriptsize PID}}$~~  &  $~1.5123$ ~&~$1.2983$~& ~$0.7757$ ~& $26-30$sec.\\ \hline
~Intelligent LQR$_{\mbox{\scriptsize  PID}}$~&~$0.9428$ & $0.9193$ & $0.2096$ & $26-30$sec. \\ \hline
\end{tabular}
\end{center}
\label{sample-tableP}
\end{small}
   \vspace{0.15cm} 
\end{table}

\begin{table}[ht]
\begin{small}
\caption{ \textbf{Yaw} tracking errors}
\begin{center}
\begin{tabular}{| c || c c  c c |}
  \hline
~Controller~  & ~~ RMS~~  &  STD  & Mean &~~~~Interval~~~~   \\ \hline
~~LQR$_{\mbox{\scriptsize PID}}$~~  &  $~6.9951$ ~&~$0.2825$~& ~$6.9893$ ~& $26-30$sec.\\ \hline
~Intelligent LQR$_{\mbox{\scriptsize  PID}}$~&~$2.3304$ & $0.1981$ & $2.3220$ & ~$26-30$sec. \\ \hline
\end{tabular}
\end{center}
\label{sample-tableY}
\end{small}
   \vspace{0.15cm} 
\end{table}

\subsection{Parameter uncertainty}\label{sec-PU}
\begin{figure*}[ht]
\vspace{0.25cm}
   \begin{minipage}[c]{0.475\linewidth}  
        \centering
      \begin{overpic}[scale=0.32]{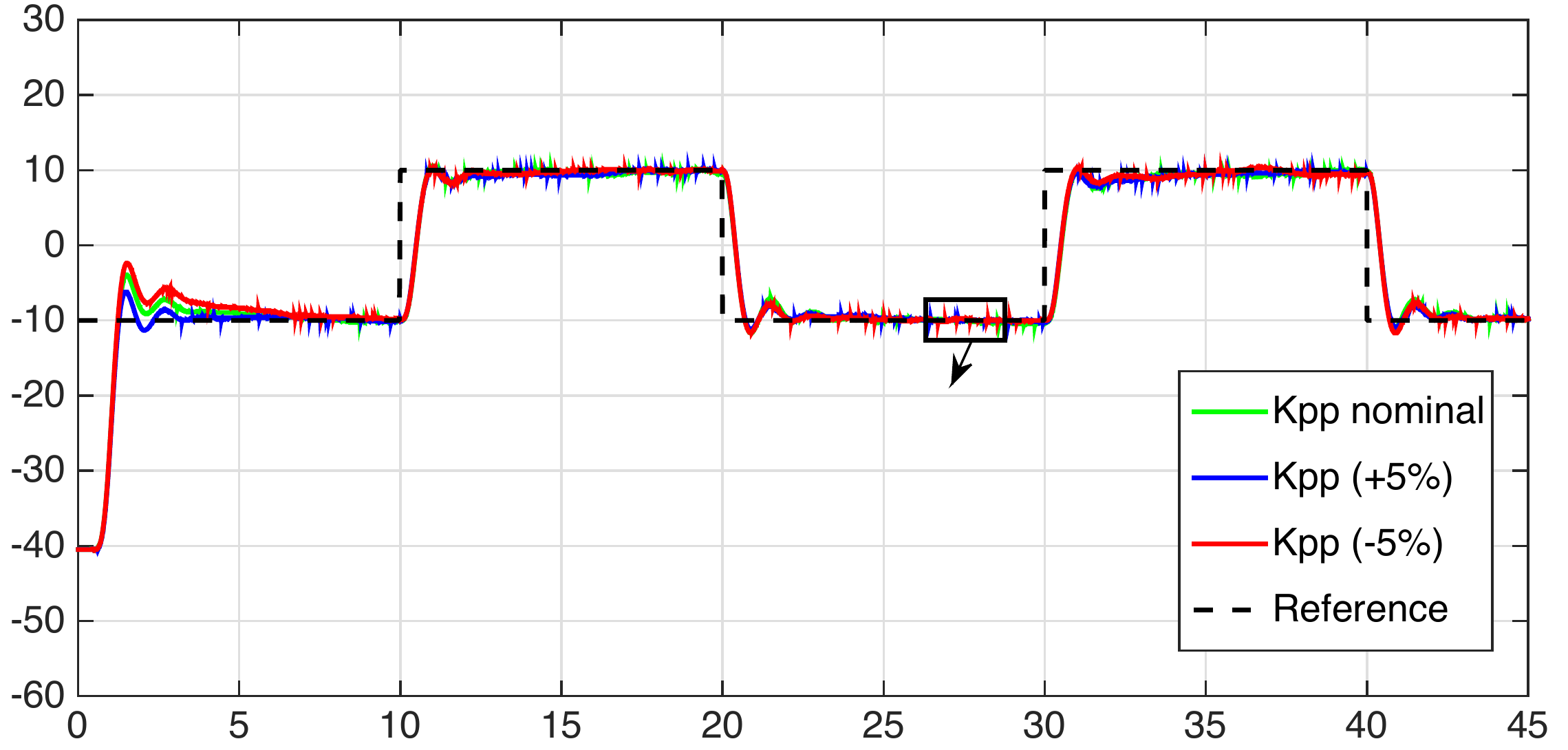}
      \put(-1.5,16){\scriptsize \begin{rotate}{90} {Pitch angle $(\deg)$}\end{rotate}}
     \put(20,5){\begin{overpic}[scale=0.135]{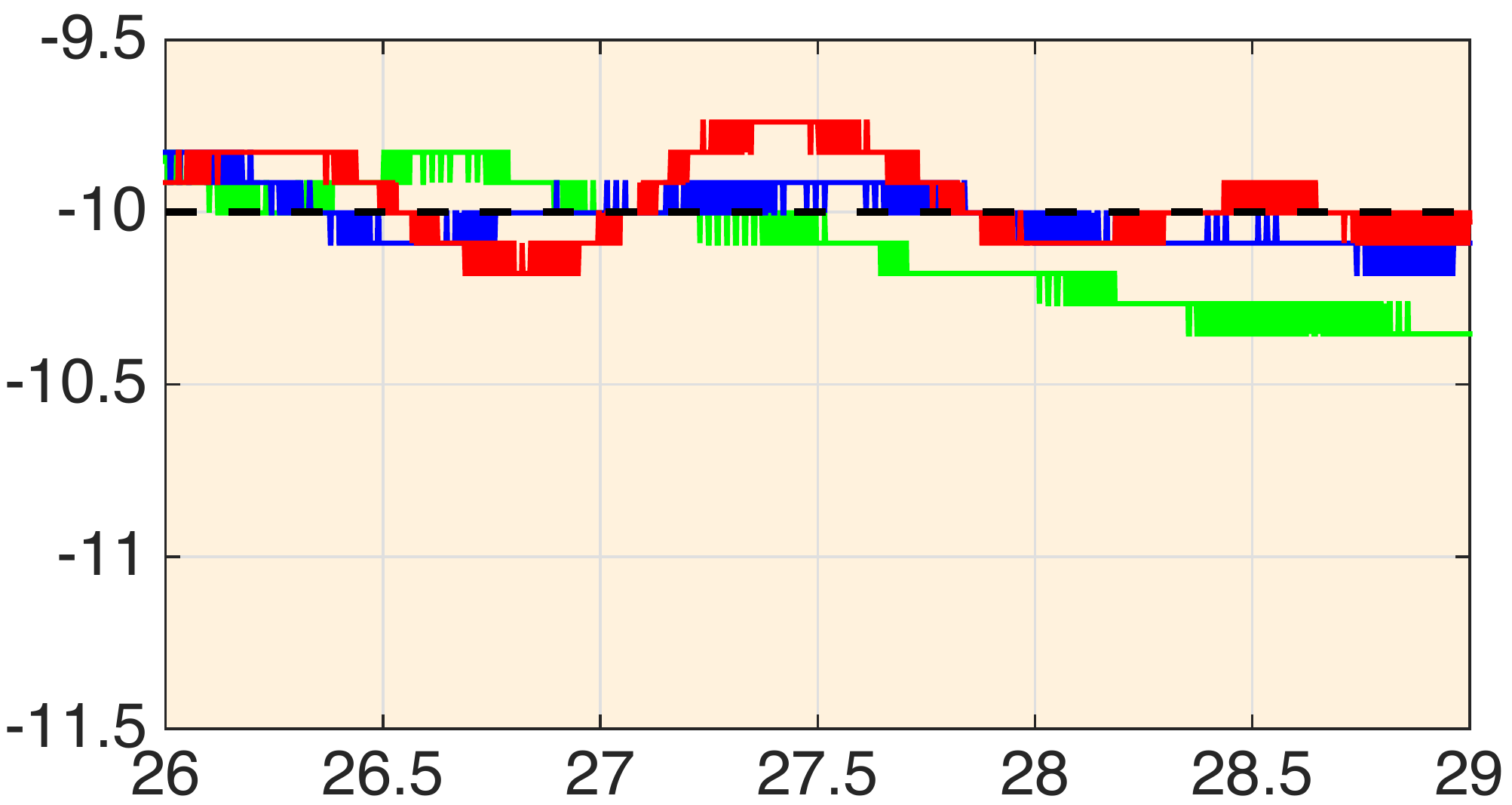}
   \end{overpic}}
 \put(42,-4){Time [sec]}
            \end{overpic}  \vspace{2pt}
              \caption{Tracking responses of pitch angle under pitch reference step during parameter uncertainty (Intelligent LQR based PID controller).}\label{fig-iPU-P}  
       \end{minipage}\hfill 
         \begin{minipage}[c]{0.475\linewidth}
      \begin{overpic}[scale=0.32]{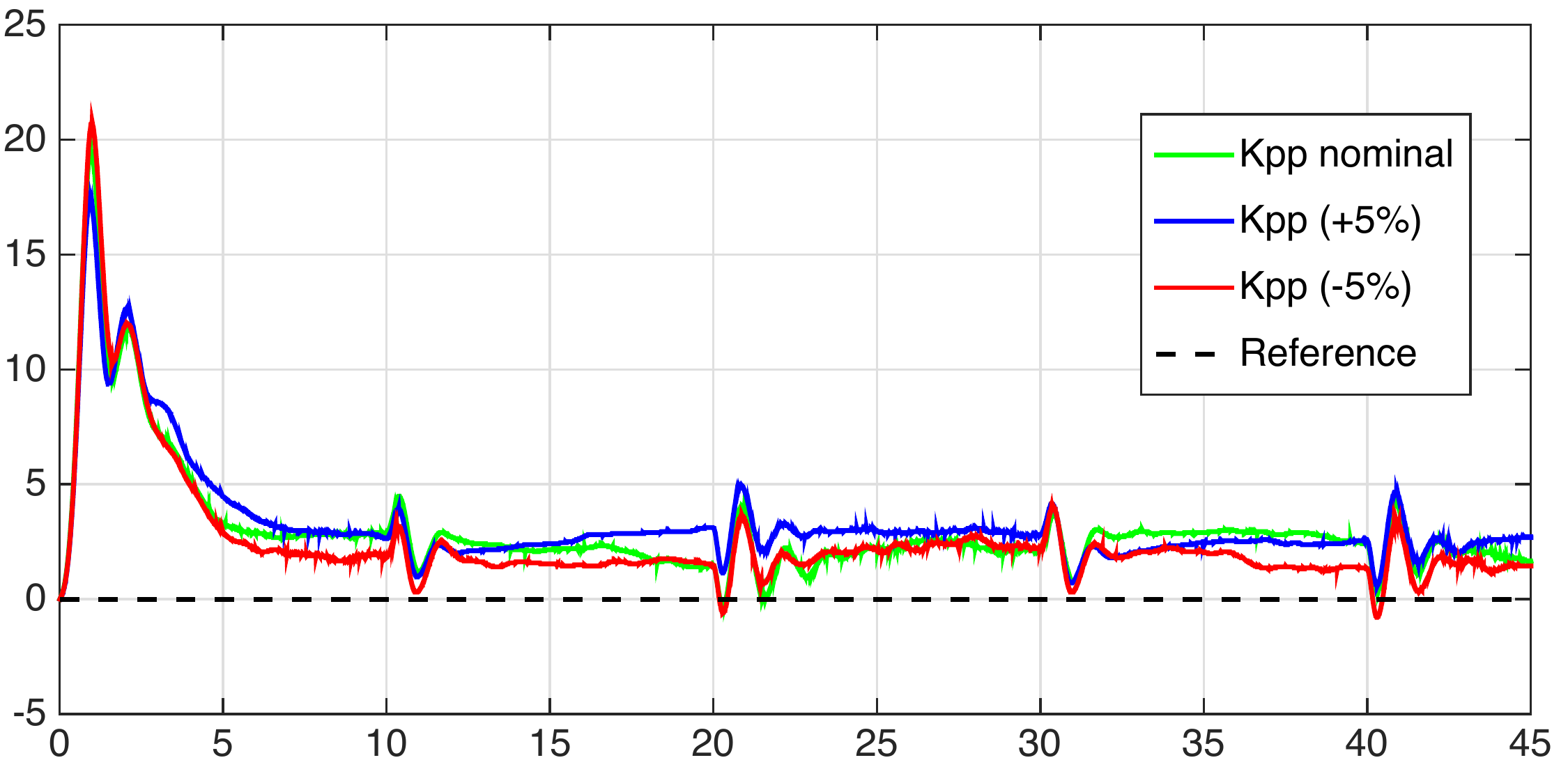}
      \put(-1.5,16){\scriptsize \begin{rotate}{90} {Yaw angle $(\deg)$}\end{rotate}}
 \put(42,-4){ Time [sec]}
            \end{overpic}  \vspace{2pt}
              \caption{Tracking responses of yaw angle under pitch reference step during parameter uncertainty (Intelligent LQR based PID controller).}\label{fig-iPU-Y}
   \end{minipage} 
   \vspace{0.25cm}     
\end{figure*}

\begin{figure*}[ht]
\vspace{0.25cm}
   \begin{minipage}[c]{0.475\linewidth}  
        \centering
      \begin{overpic}[scale=0.32]{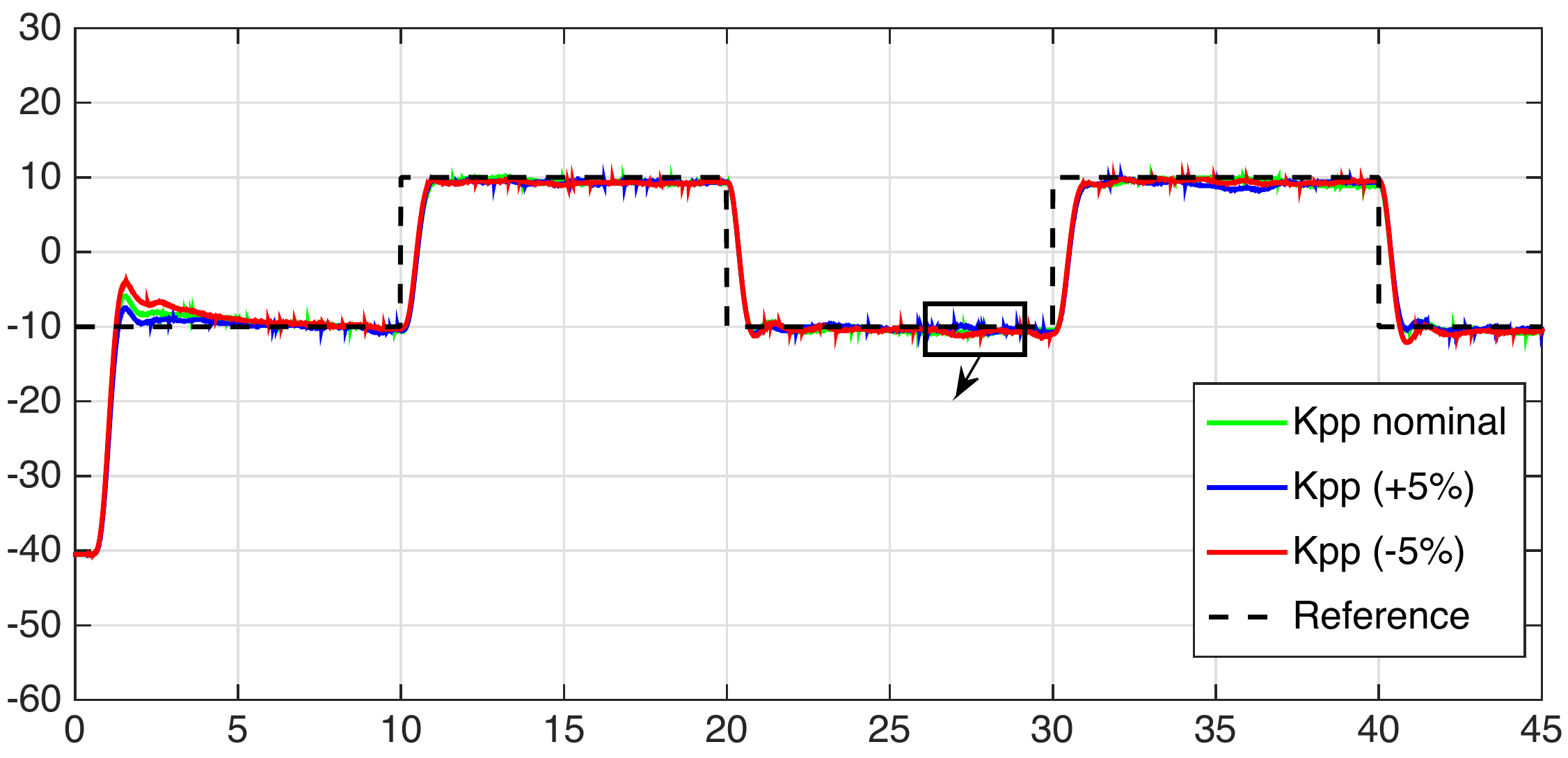}
      \put(-1.5,16){\scriptsize \begin{rotate}{90} {Pitch angle $(\deg)$}\end{rotate}}
     \put(20,5){\begin{overpic}[scale=0.13]{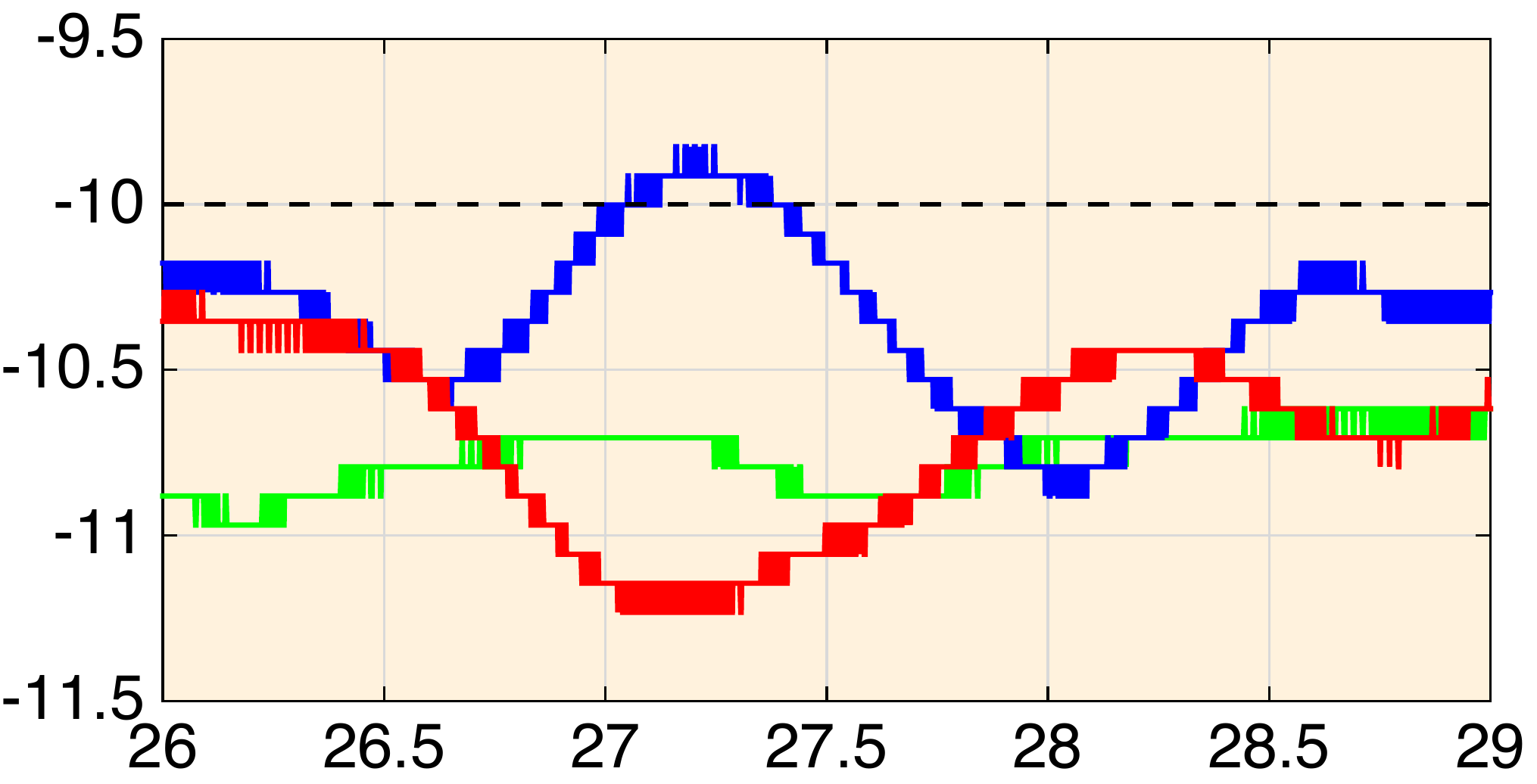}
   \end{overpic}}
 \put(42,-4){Time [sec]}
            \end{overpic}  \vspace{1pt}
        \caption{Tracking responses of pitch angle under pitch reference step during parameter uncertainty (LQR based PID controller).}\label{fig-PU-P}
       \end{minipage}\hfill 
         \begin{minipage}[c]{0.475\linewidth}
      \begin{overpic}[scale=0.32]{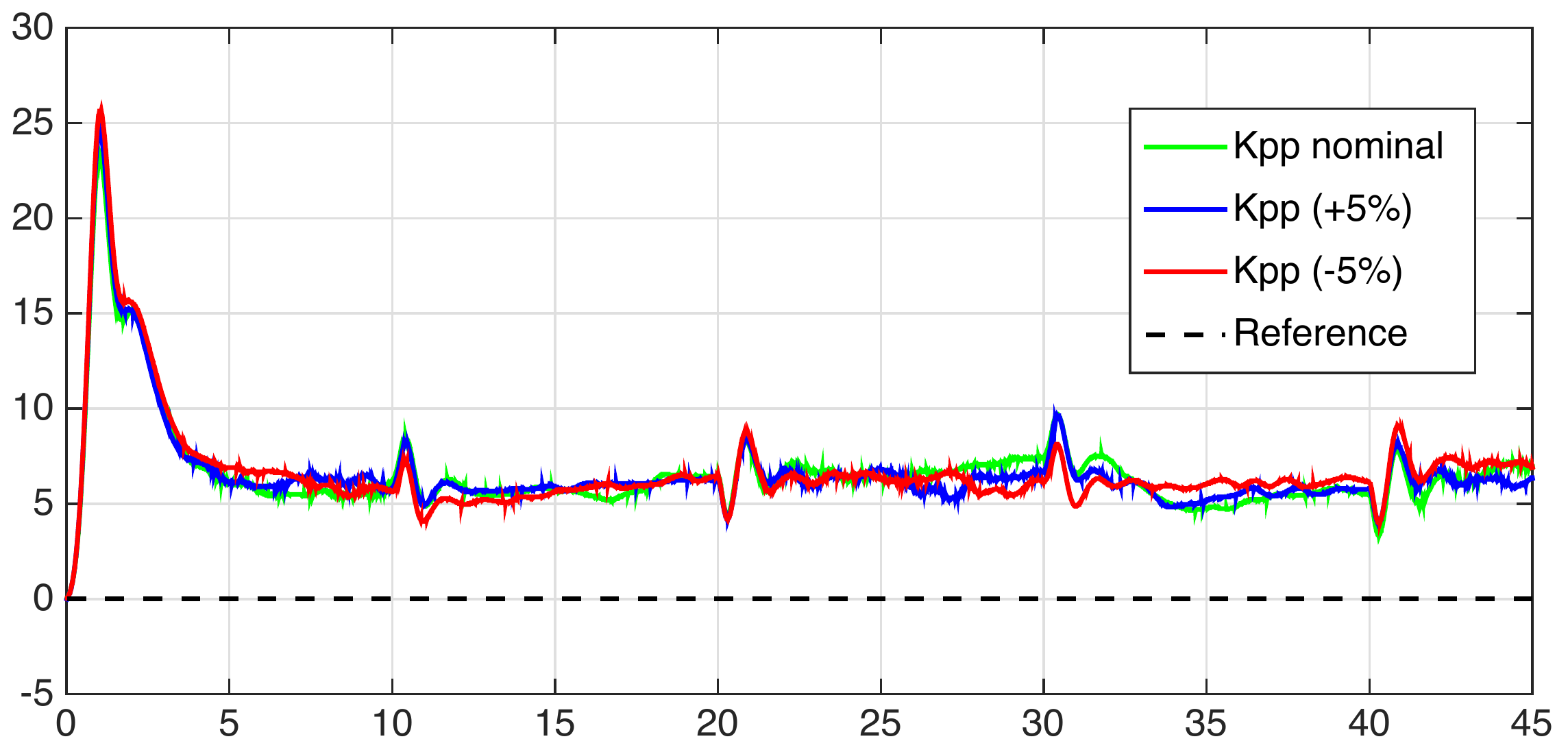}
      \put(-1.5,16){\scriptsize \begin{rotate}{90} {Yaw angle $(\deg)$}\end{rotate}}
 \put(42,-4){Time [sec]}
            \end{overpic}  \vspace{1pt}
        \caption{Tracking responses of yaw angle under pitch reference step during parameter uncertainty (LQR based PID controller).}\label{fig-PU-Y}
   \end{minipage} 
   \vspace{0.25cm}     
\end{figure*}

The robustness of the control scheme against model variation is assessed in this subsection.  Model uncertainty is introduced at the yaw angle. Two test cases  are considered where  $+5\%$ and $-5\%$ of $K_{pp}$ pitch force constant parameter variations have been introduced respectively. Figs. \ref{fig-iPU-P} and \ref{fig-iPU-Y} show both pitch and yaw responses of i-LQR-PID and Figs. \ref{fig-PU-P} and \ref{fig-PU-Y} illustrate both pitch and yaw responses of LQR-PID controller respectively for both levels of uncertainties. The level of the oscillations in case of $5\%$ of variations  is higher  when using  i-LQR-PID controller. Furthermore, the convergence time of yaw angles for i-LQR-PID controller is faster and the maximum offset is smaller and much better than LQR-PID controller response.

\subsection{Disturbance rejection}\label{sec-DR}
\begin{figure*}[ht]
\vspace{0.25cm}
   \begin{minipage}[c]{0.475\linewidth}  
        \centering
      \begin{overpic}[scale=0.34]{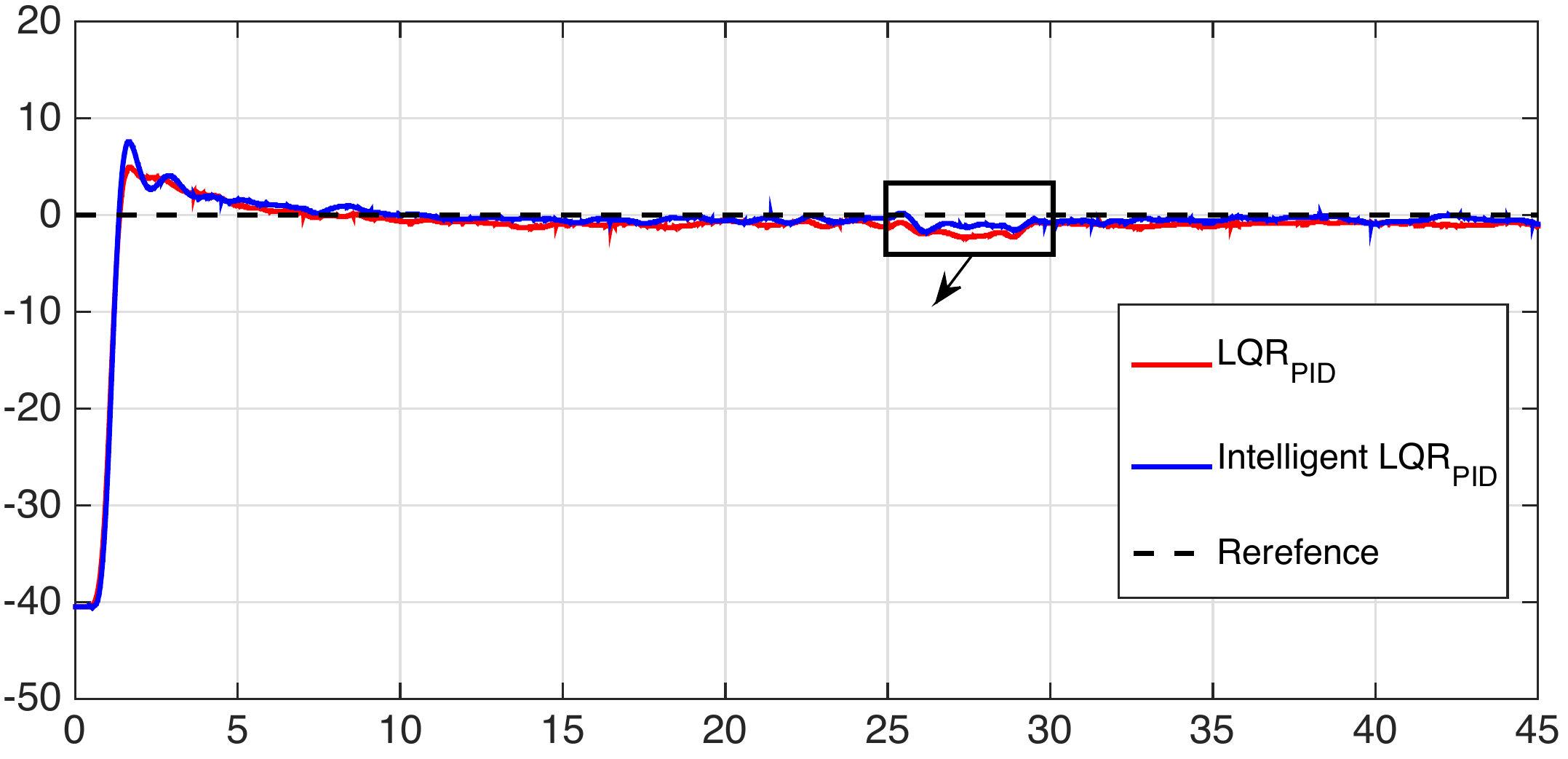}
      \put(-1.5,16){\scriptsize \begin{rotate}{90} {Pitch angle $(\deg)$}\end{rotate}}
     \put(22,6){\begin{overpic}[scale=0.13]{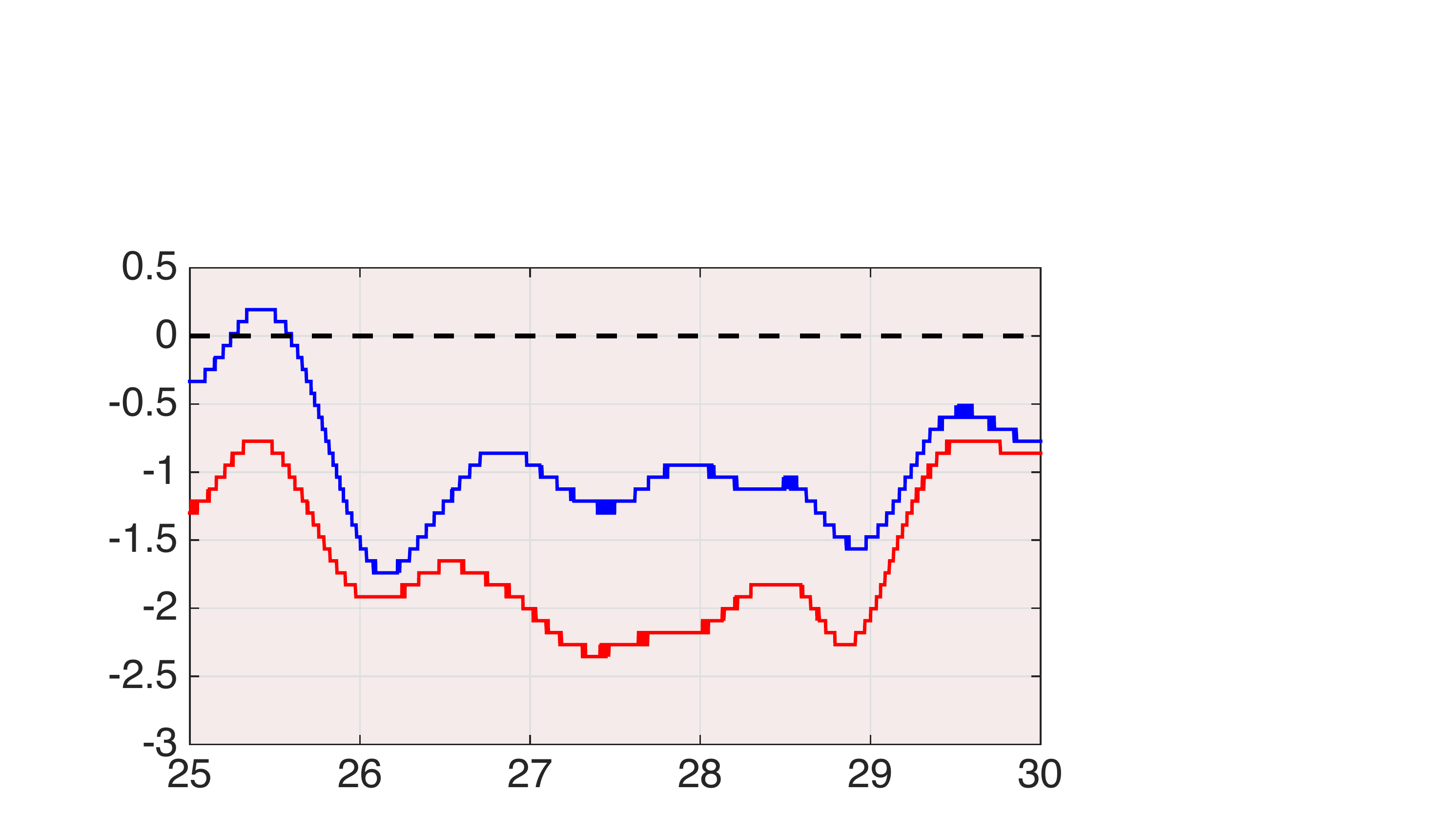}
   \end{overpic}}
 \put(42,-4){ Time [sec]}
            \end{overpic}  \vspace{2pt}
              \caption{Tracking responses of pitch angle during short term disturbance.}\label{fig-SD-P}
       \end{minipage}\hfill 
         \begin{minipage}[c]{0.475\linewidth}
      \begin{overpic}[scale=0.34]{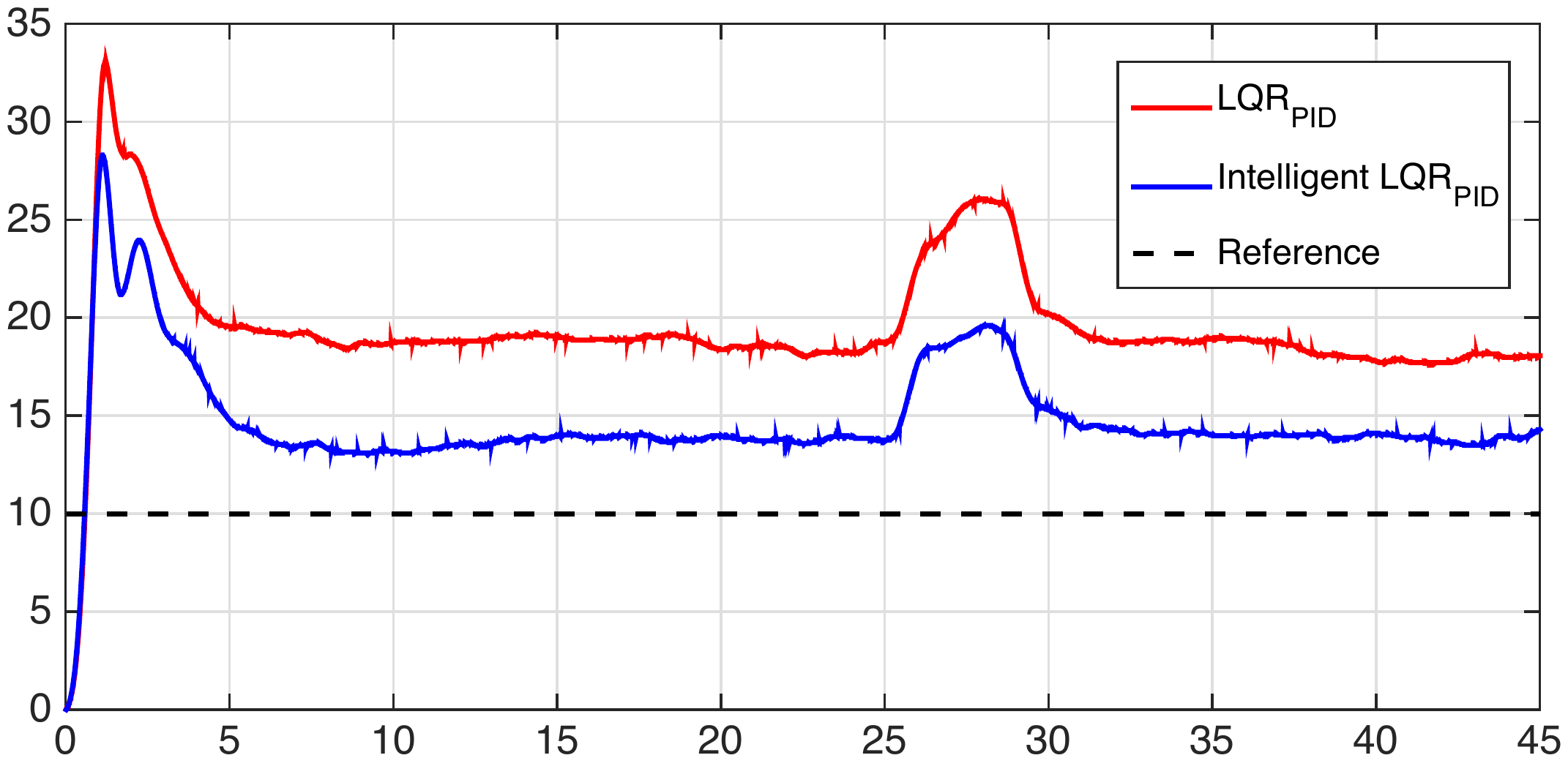}
      \put(-1.5,16){\scriptsize \begin{rotate}{90} {Yaw angle $(\deg)$}\end{rotate}}
 \put(42,-4){Time [sec]}
            \end{overpic}  \vspace{2pt}
              \caption{Tracking responses of yaw angle during short term disturbance.}\label{fig-SD-Y}
   \end{minipage} 
   \vspace{0.25cm}     
\end{figure*}

The robustness of i-LQR-PID is  assessed in  two test cases where  short term disturbance and continuous disturbance are respectively introduced into the 2-DoF helicopter system.
\begin{itemize}
\item {\it Short term disturbance}
\end{itemize}
The short term bounded pulse external disturbance function which $10^{\circ}$ disturbance magnitude, $35$seconds period, $25$seconds phase delay and $10\%$ pulse width is introduced into the yaw control angle from the time interval $[25-30]$seconds. Figs. \ref{fig-SD-P} and \ref{fig-SD-Y}, which illustrate the pitch and yaw responses during a short term disturbance experiment for both i-LQR-PID and LQR-PID controllers, demonstrate the ability of both controllers to reduce the deviation and brings back the response to set-point in around $t\!=\!5$ seconds. Moreover, i-LQR-PID provides better satisfactory results in reducing the pitch and yaw amplitudes in shorter time. 

\begin{figure*}[ht]
\vspace{0.25cm}
   \begin{minipage}[c]{0.475\linewidth}  
        \centering
      \begin{overpic}[scale=0.34]{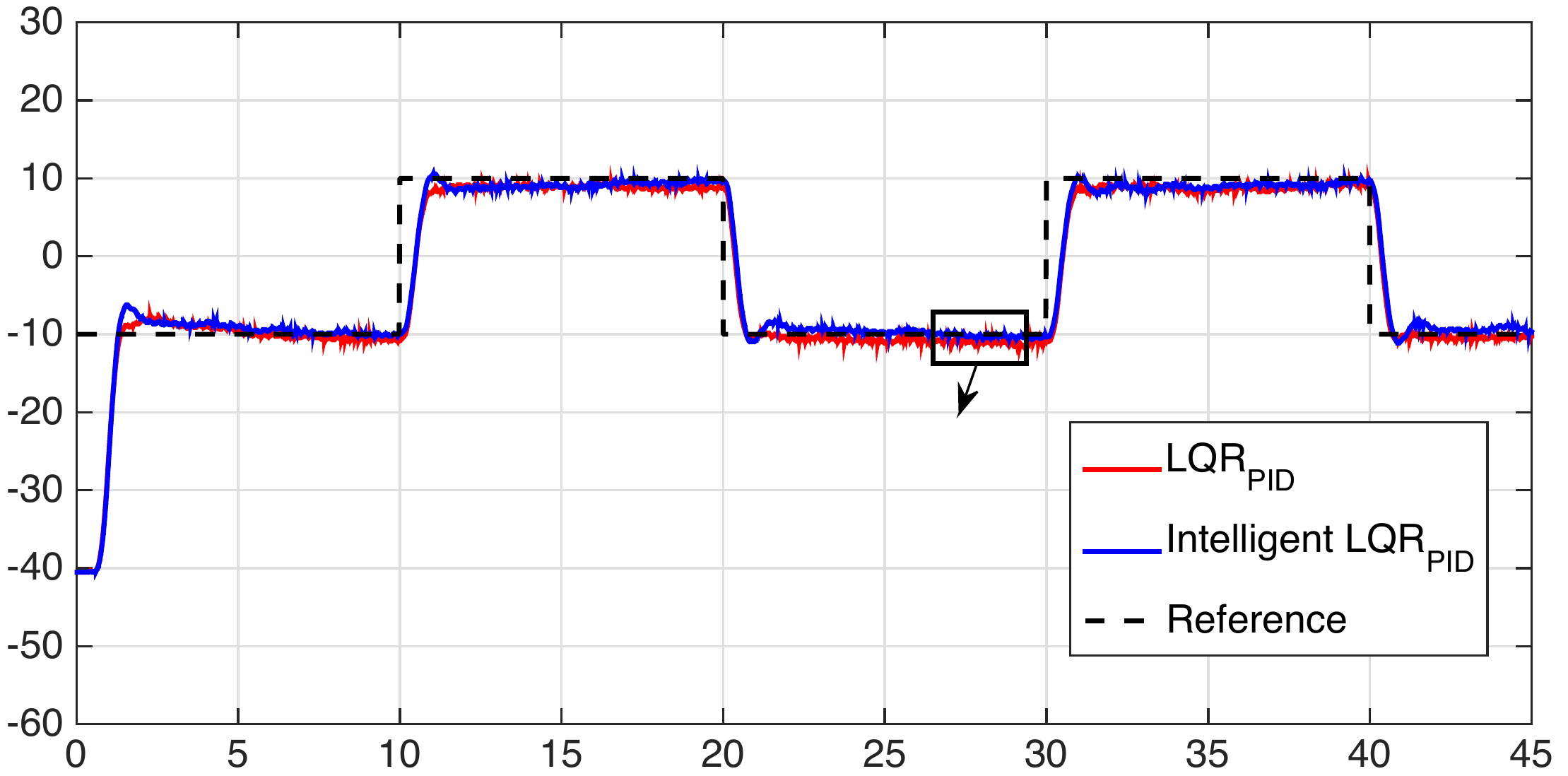}
      \put(-1.5,16){\scriptsize \begin{rotate}{90} {Pitch angle $(\deg)$}\end{rotate}}
     \put(22,5){\begin{overpic}[scale=0.13]{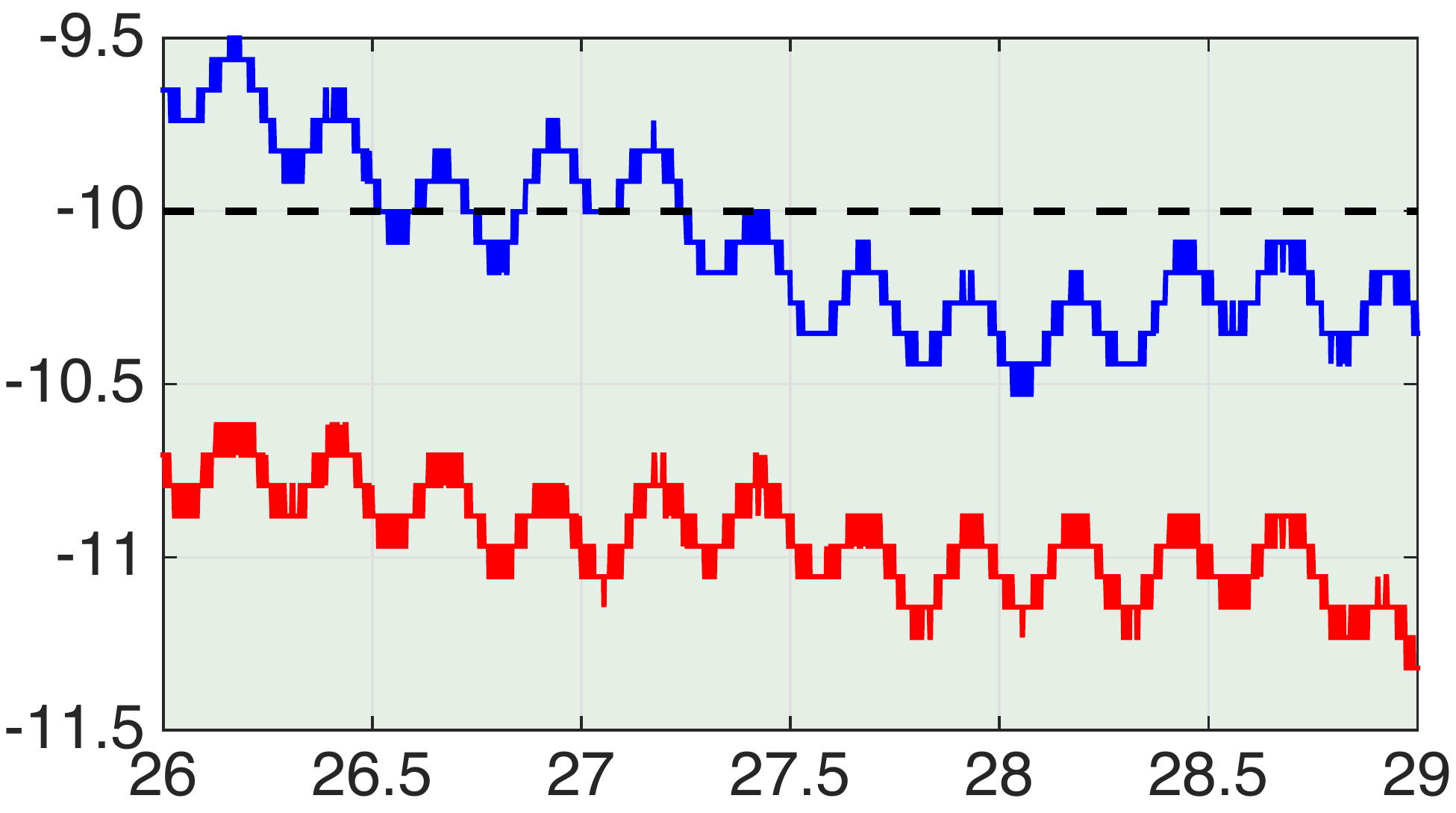}
   \end{overpic}}
 \put(42,-4){Time [sec]}
            \end{overpic}  \vspace{2pt}
              \caption{Pitch tracking responses under pitch reference step during continuous disturbance.}\label{fig-CD-P}
       \end{minipage}\hfill 
         \begin{minipage}[c]{0.475\linewidth}
      \begin{overpic}[scale=0.34]{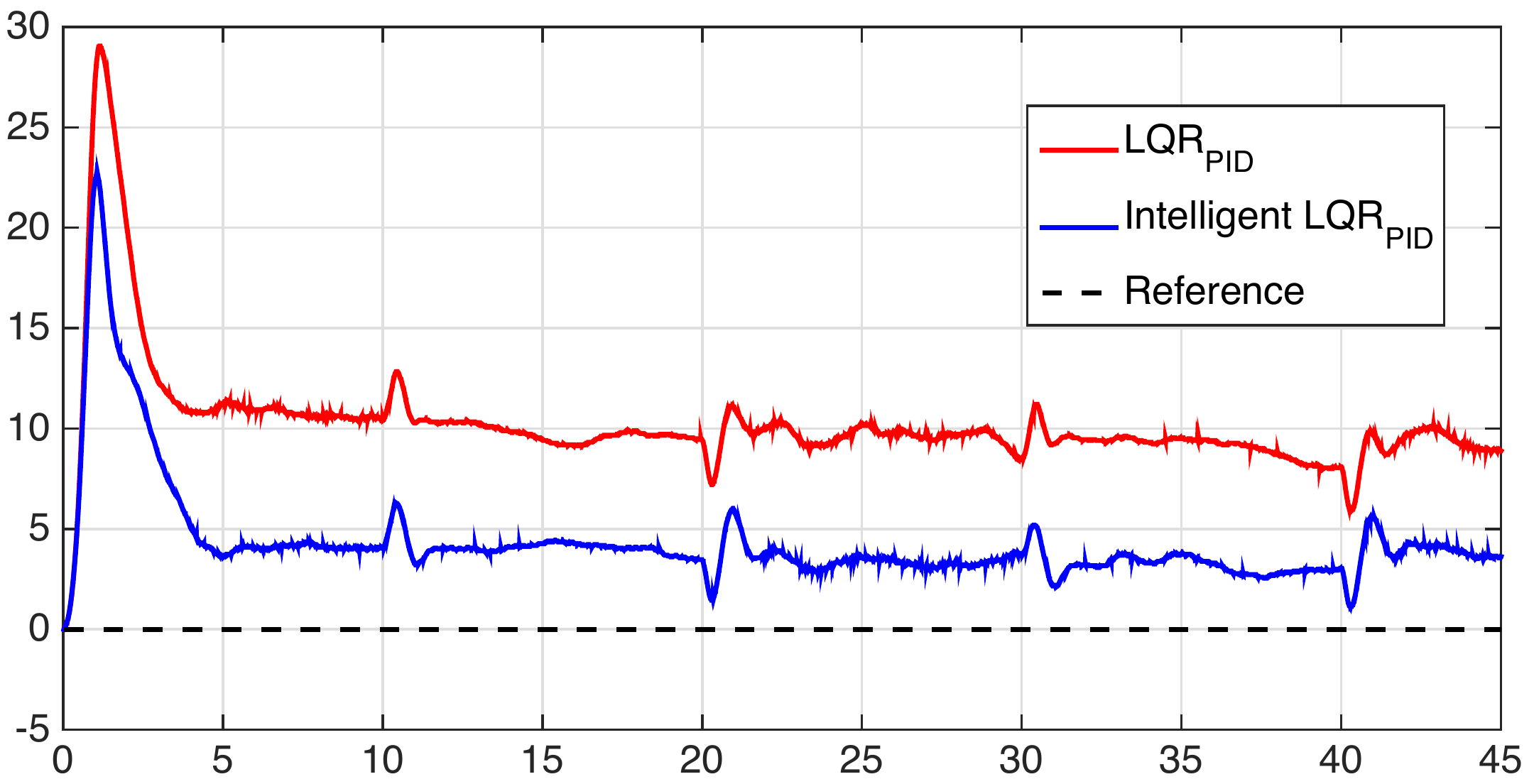}
      \put(-1.5,16){\scriptsize \begin{rotate}{90} {Yaw angle $(\deg)$}\end{rotate}}
 \put(42,-4){Time [sec]}
            \end{overpic}  \vspace{2pt}
              \caption{Yaw tracking responses under pitch reference step during continuous disturbance.}\label{fig-CD-Y}
   \end{minipage} 
   \vspace{0.25cm}     
\end{figure*}

\begin{itemize}
  \item {\it Continuous disturbance}
\end{itemize}
The continuous disturbance function $10\sin(25t+10)$ which $10^{\circ}$ disturbance magnitude is introduced during the step command tracking of pitch control angle 2-DoF helicopter. Figs. \ref{fig-CD-P} and \ref{fig-CD-Y} show the response of both i-LQR-PID and LQR-PID controllers framework during the continuous disturbance.
It can be noted that the deviation in magnitude is restricted within $\pm0.3^{\circ}$ for both i-LQR-PID and LQR-PID controllers. The ability of both controllers to reject the disturbance and track the reference signal is highlighted in the zoomed view of the pitch response shown in Fig. \ref{fig-CD-P}.

\subsection{Robustness under wind gusts disturbances}\label{sec-real-time}
\begin{figure*}[ht]
\vspace{0.25cm}
   \begin{minipage}[c]{0.475\linewidth}  
        \centering
      \begin{overpic}[scale=0.34]{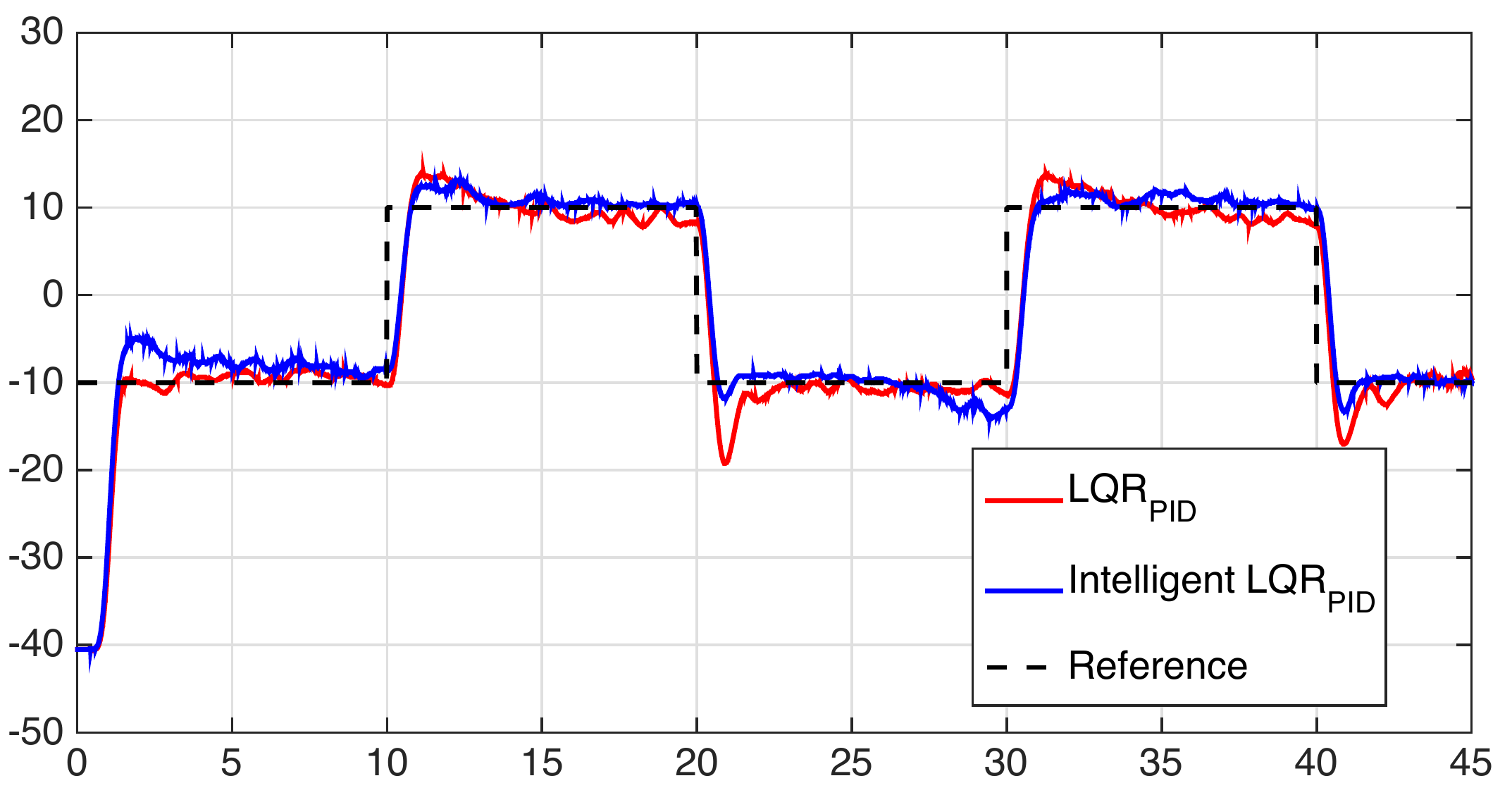}
      \put(-1.5,16){\scriptsize \begin{rotate}{90} {Pitch angle $(\deg)$}\end{rotate}}
 \put(42,-4){Time [sec]}
            \end{overpic}  \vspace{2pt}
              \caption{Pitch tracking responses under pitch reference step during continuous wind disturbance.}\label{fig-windP}
       \end{minipage}\hfill 
         \begin{minipage}[c]{0.475\linewidth}
      \begin{overpic}[scale=0.34]{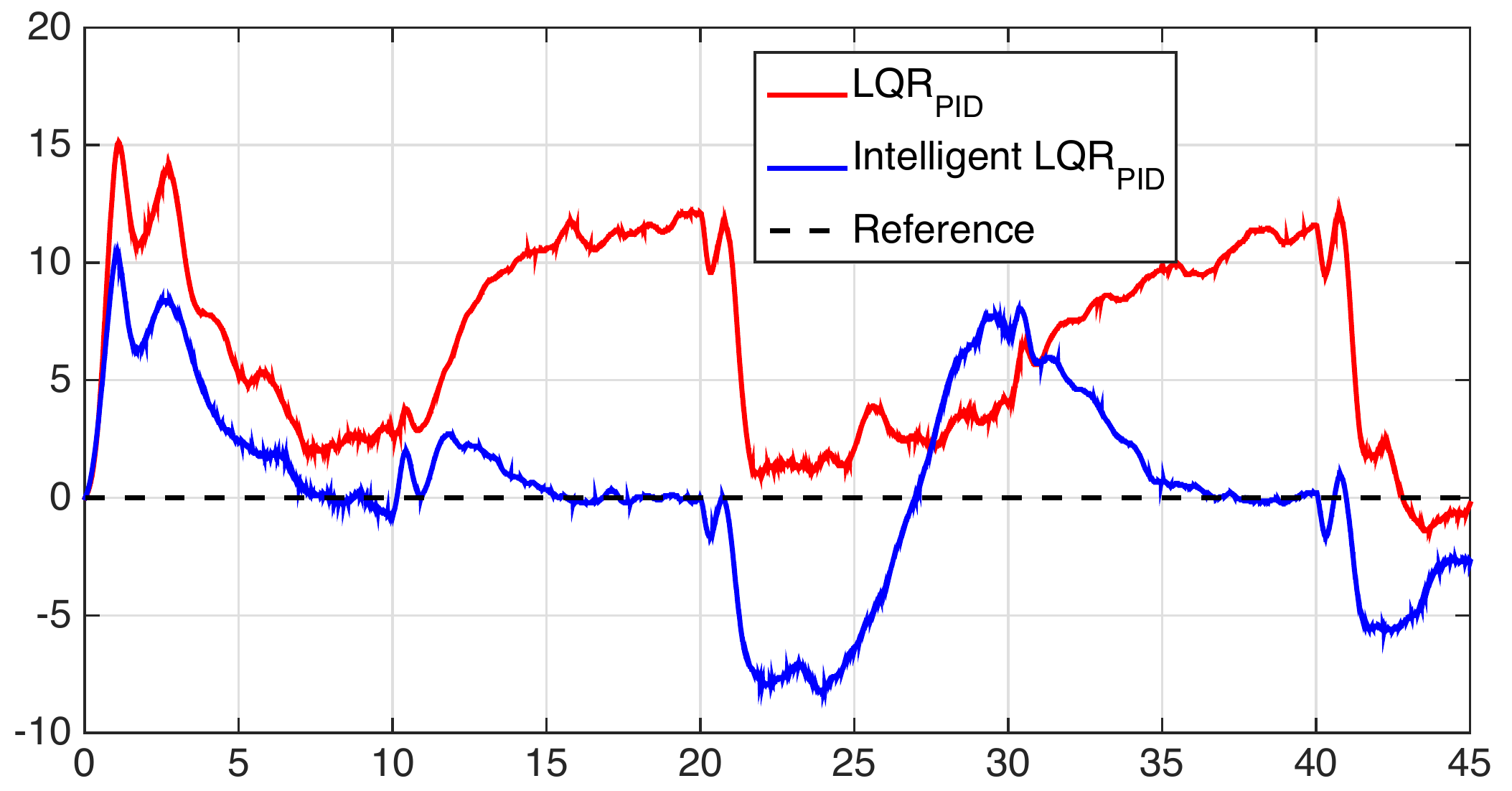}
      \put(-1.5,16){\scriptsize \begin{rotate}{90} {Yaw angle $(\deg)$}\end{rotate}}
 \put(42,-4){Time [sec]}
            \end{overpic}  \vspace{2pt}
              \caption{Yaw tracking responses under pitch reference step during continuous wind disturbance.}\label{fig-windY}
   \end{minipage} 
   \vspace{0.25cm}     
\end{figure*}

In real time scenarios,  in addition to varying magnitudes  the helicopter needs often  to adapt  to directional changes. Hence, position tracking control problem under aggressive wind turbulence effects is investigated. The wind gusts is generated by an electrical fan and with  fixed velocity. The wind velocity parallels the pitch axis when the pitch and yaw angles are set to $\theta\!=\!-40.5^{\circ}$ and $\Psi\!=\!0^{\circ}$ respectively. Figs. \ref{fig-windP} and \ref{fig-windY} show the maneuvering performance of both i-LQR-PID and LQR-PID controllers for pitch and yaw angles. The i-LQR-PID controller provides  better trajectory tracking results.

\begin{figure*}[ht]
        \centering
      \begin{overpic}[scale=0.13]{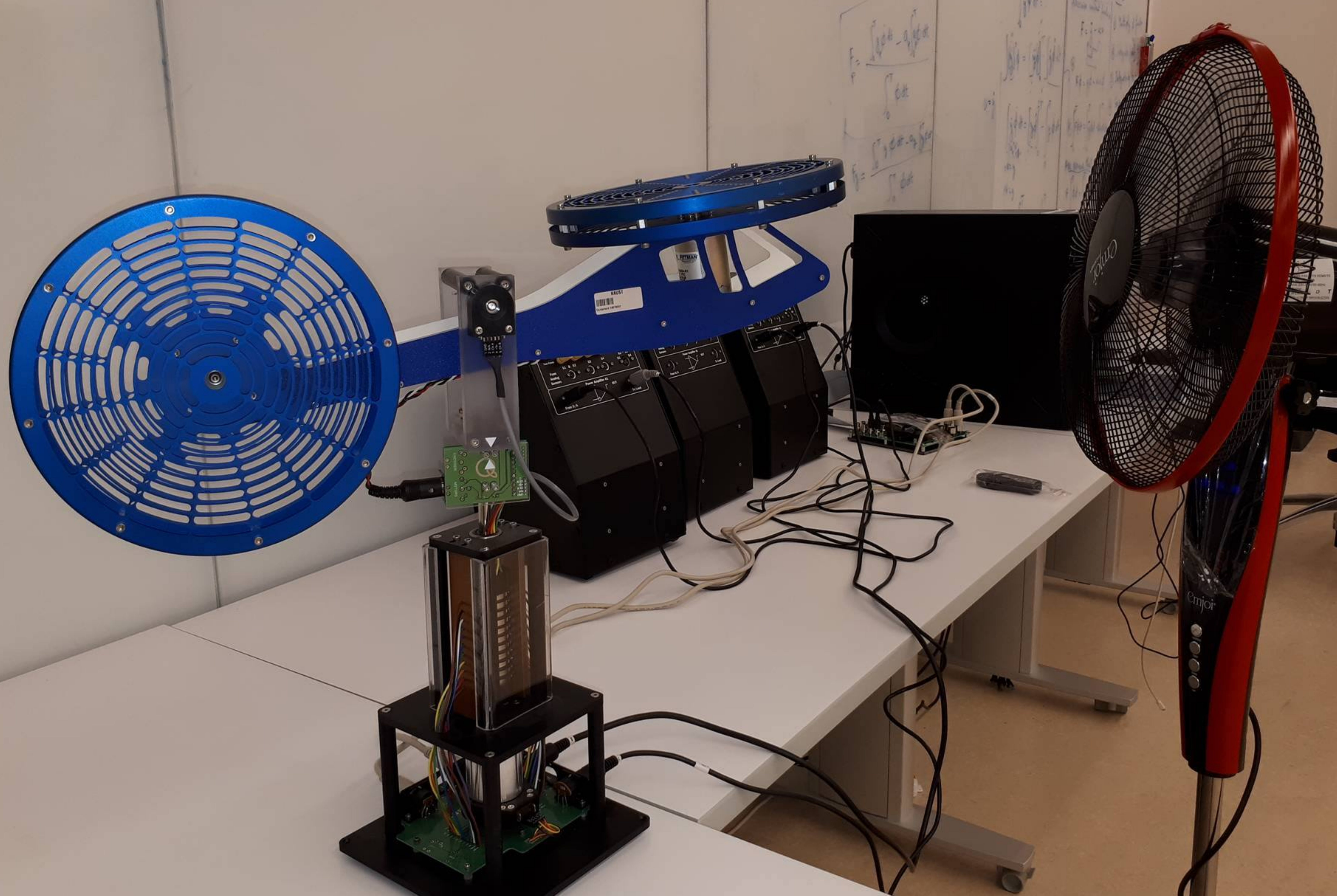}
            \end{overpic}  \vspace{2pt}
              \caption{Experimental setup under wind gust.}\label{fig-wind}
\end{figure*}  
            
\begin{table}[ht]
\vskip 0.15in
\begin{small}
\caption{RMS tracking errors of \textbf{Pitch} and \textbf{Yaw} angles under wind disturbances}
\begin{center}
\begin{tabular}{| c || c || c || c |}
  \hline
~Controller~  & ~~ RMS (pitch)~~  &  ~~RMS (yaw)~~  & ~Interval~~~~   \\ \hline
~~LQR$_{\mbox{\scriptsize PID}}$~~  &  $5.8598$ ~&~$7.7559$~&$0-45$sec.\\ \hline
~Intelligent LQR$_{\mbox{\scriptsize  PID}}$~&$5.8673$ & $4.0847$ &  $0-45$sec. \\ \hline
\end{tabular}
\end{center}
\label{sample-tablePY}
\end{small}
   \vspace{0.15cm} 
\end{table}

The performance indices are given in Table \ref{sample-tablePY} for both pitch and yaw angles respectively. From Table \ref{sample-tablePY}, it can be clearly seen that i-LQR-PID controller achieves better trajectory tracking performances than LQR-PID.

\newpage

\section{Conclusion}\label{conclusion}
In this paper, i-LQR-PID controller has been proposed to control the pitch and yaw angles so as to make the system to track the reference trajectory. The performance of the presented i-LQR-PID has been evaluated and compared to the LQR-PID controller in closed-loop to accommodate the disturbances present in the 2-DoF helicopter system. Simulation and experimental results of 2-DoF helicopter for different level of magnitudes and direction have shown that i-LQR-PID controller is more effective and robust than LQR-PID controller for tracking references under aggressive turbulence effects, while it preserves its simplicity of implementation. 



\bibliography{biblio.bib}

\end{document}